\documentclass[epj,nopacs]{svjour}

\usepackage[breaklinks=true,colorlinks=true,linkcolor=blue,urlcolor=blue,citecolor=blue]{hyperref}

\usepackage[pdftex]{graphicx}
\usepackage{xcolor}
\usepackage{siunitx}
\usepackage[T1]{fontenc}
\usepackage{lmodern}
\usepackage{amsmath,amsfonts}
\usepackage{lineno}
\usepackage[title]{appendix}

\usepackage[square,numbers]{natbib}
\usepackage{xurl}

\def\andname{\hspace*{-0.5em}}

\begin{document}

\title{QUEST-DMC superfluid $^3$He detector for sub-GeV dark matter}

% Authors
\author{
\textbf{QUEST-DMC~collaboration}:~S.~Autti\inst{1}\thanks{\href{mailto:s.autti@lancaster.ac.uk}{s.autti@lancaster.ac.uk}} \and
A.~Casey\inst{2} \and
N.~Eng\inst{2} \and
N.~Darvishi\inst{2} \and
P.~Franchini\inst{1}\inst{2}\thanks{\href{mailto:paolo.franchini@rhul.ac.uk}{paolo.franchini@rhul.ac.uk}} \and
R.~P.~Haley\inst{1} \and
P.J.~Heikkinen\inst{2} \and
A.~Jennings\inst{3} \and
A.~Kemp\inst{2} \and
E.~Leason\inst{2}\thanks{\href{mailto:elizabeth.leason@rhul.ac.uk}{elizabeth.leason@rhul.ac.uk}}
\and
L.V.~Levitin\inst{2} \and
J.~Monroe\inst{2} \and
J.~March-Russel\inst{4} \and
M.~T.~Noble\inst{1} \and
J.~R.~Prance\inst{1} \and
X.~Rojas\inst{2} \and
T.~Salmon\inst{1} \and
J.~Saunders\inst{2} \and
R.~Smith\inst{2} \and
M.~D.~Thompson\inst{1} \and
V.~Tsepelin\inst{1} \and
S.~M.~West\inst{2} \and
L.~Whitehead\inst{1} \and
V.~V.~Zavjalov\inst{1} \and
D.~E.~Zmeev\inst{1}
}

\institute{Department of Physics, Lancaster University, Lancaster, LA1 4YB, UK. \and
Department of Physics, Royal Holloway University of London, Egham, Surrey, TW20 0EX, UK. \and
RIKEN Center for Quantum Computing, RIKEN, Wako, 351-0198, Japan. \and
Rudolf Peierls Centre for Theoretical Physics, University of Oxford 1 Keble Road, Oxford, OX1 3NP, UK.}

\date{Received: 20 July 2023 / Revised version:  5 January 2024}

\abstract{
The focus of dark matter searches to date has been on Weakly Interacting Massive Particles (WIMPs) in the GeV/$c^2$-TeV/$c^2$ mass range. The direct, indirect and collider searches in this mass range have been extensive but ultimately unsuccessful, providing a strong motivation for widening the search outside this range. Here we describe a new concept for a dark matter experiment, employing superfluid 
$^3$He as a detector for dark matter that is close to the mass of the proton, of order 1\,GeV/$c^2$. 
The QUEST-DMC detector concept is based on quasiparticle detection in a bolometer cell by a nanomechanical resonator. In this paper we develop the energy measurement methodology and detector response model, simulate candidate dark matter signals and expected background interactions, and calculate the sensitivity of such a detector. We project that such a detector can reach sub-eV nuclear recoil energy threshold, opening up new windows on the parameter space of both spin-dependent and spin-independent interactions of light dark matter candidates.
}

\maketitle
%\begin{figure}[!htb]
%\includegraphics[width=1\linewidth]{.pdf}%
%\caption{\label{fig:demag2}}
%\end{figure}

\section{Introduction} % All

Dark matter plays a vital role in the evolution of the universe, for example, it played a central role in the formation of structure in the early universe and today plays a key role in stopping galaxies flying apart. The focus of dark matter studies and searches to date has been on Weakly Interacting Massive Particles (WIMPs) whose predicted mass range is broadly speaking between 10-1000 times that of the proton. The direct, indirect and collider searches for this dark matter candidate to date have been extensive but so far unsuccessful. There is a strong motivation to widen the search.

% Typically direct detection searches focus on detection elastic dark matter - nucleus scattering. However, for low dark matter masses (below 1 GeV) both the kinetic energy and transferred recoil energy fall rapidly. This motivates the need for experiments with very low energy thresholds. Additionally the recoil energy for sub-GeV dark matter is inversely proportional to the mass of the target nucleus motivating the use of a low mass target nucleus.

Dark matter with a mass of order or below the mass of the proton, commonly referred to as the ``sub-GeV/$c^2$'' mass region has received significant recent attention driven by the widening exploration of possible connections between the visible and hidden sectors. Coupled with the idea that hidden sectors may contain more than just the dark matter state leads to a rich set of new dark matter-Standard Model interactions with novel dark matter genesis mechanisms moving beyond the vanilla WIMP picture. A comprehensive list of sub-GeV/$c^2$ models with a review of the growing international landscape of small detectors for low-mass dark matter searches can be found in~\cite{Essig:2022dfa}.

%~\cite{Nussinov:1985xr,Kaplan:1991ah,Farrar:2004qy,Hooper:2004dc,Kaplan:2009ag,Falkowski:2011xh,Feng:2008ya,Feng:2011ik, McDonald:2001vt,Hall:2009bx,Essig:2011nj,Boehm:2003ha,Pospelov:2008jk,Hochberg:2014dra,Smirnov:2020zwf}. 

% The fact that the universe only consists of matter with no anti-matter requires explanation, since it is reasonable to assume that matter and anti-matter were produced in equal quantities in the Big Bang. This implies that during the evolution of the universe a process took place that dynamically generated the asymmetry between matter and anti-matter. Models linking the dynamics of dark matter with the generation of the matter/anti-matter asymmetry naturally predict a mass scale of dark matter that is close to the mass of the proton, of order $\SI{1}{\giga\electronvolt \per c^2}$, suggesting an alternative target mass range to the standard WIMP. 

Helium is an attractive target for scattering searches as its relatively light mass is well-matched, kinematically, 
for sensitivity to dark matter elastically scattering off nuclei in the GeV mass range.
%the more closely matched the projectile dark matter and target atom are, the greater the momentum transfer in the interaction for the same dark matter mass.  
As a noble element, helium has the ease of purification~\cite{dmitriev2005separation}, high light yield, and transparency to its own scintillation produced in excimer decay.
%\cite{kolb2018basic} [1] \dz{maybe this?} %\smw{check whether useful}. \sautti{Dark matter %people, please fix this reference.}

Helium-4 based dark matter detection, to search for sub-GeV/$c^2$ dark matter candidates with spin-independent interactions, is being intensively explored, e.g. by the HeRALD and DELight collaborations~\cite{HeRALD2019,DELight2022}, who both employ the noble gas as their target media. 
%, aiming at a recoil energy threshold of ~1~eV~\cite{HeRALD2019,DELight2022}. 
The QUEST-DMC programme proposes to build and demonstrate the capability of a superfluid helium-3 detector to improve on the sensitivity to spin-dependent dark matter in a unique and complimentary search in the same well-motivated sub-GeV/$c^2$ dark matter mass range.
%, aiming for 1\,eV, recoil energy threshold. 
% SMW changed 10 ev to 1 ev - please note in case I have misunderstood something. 
%QUEST-DMC aims to create and operate a detector for the direct search of dark matter with sub-GeV masses using superfluid helium-3 as a target.
%aiming for sensitivity reaching 10$^{-32}$ cm$^2$ on the spin-dependent interaction cross section at 0.5 GeV dark matter mass. 
The use of superfluid $^3$He as a particle detector was first proposed 
%by G.\,R.~Pickett 
~\cite{Pickett_1988} 
in 1988. Later, {MACHe3} project led to the {ULTIMA} programme~\cite{bib:Mayet:2000aa, bib:Winkelmann:2007}, exploring a 100\,GeV/$c^2$ WIMP dark matter detector. The second quantum revolution has led to very significant improvements in achievable energy resolution since then, and QUEST-DMC aims to exploit these to reach world-leading sensitivity to spin-dependent interactions of sub-GeV/$c^2$ mass dark matter.
%is deploying SQUID readout of a nanomechanical resonator (NEMS) in the superfluid macroscopic quantum state.
%potentially enabling world-leading sensitivity to spin-dependent interactions of sub-GeV mass dark matter.

\begin{figure}[b!]
    \includegraphics[width=0.8\linewidth]{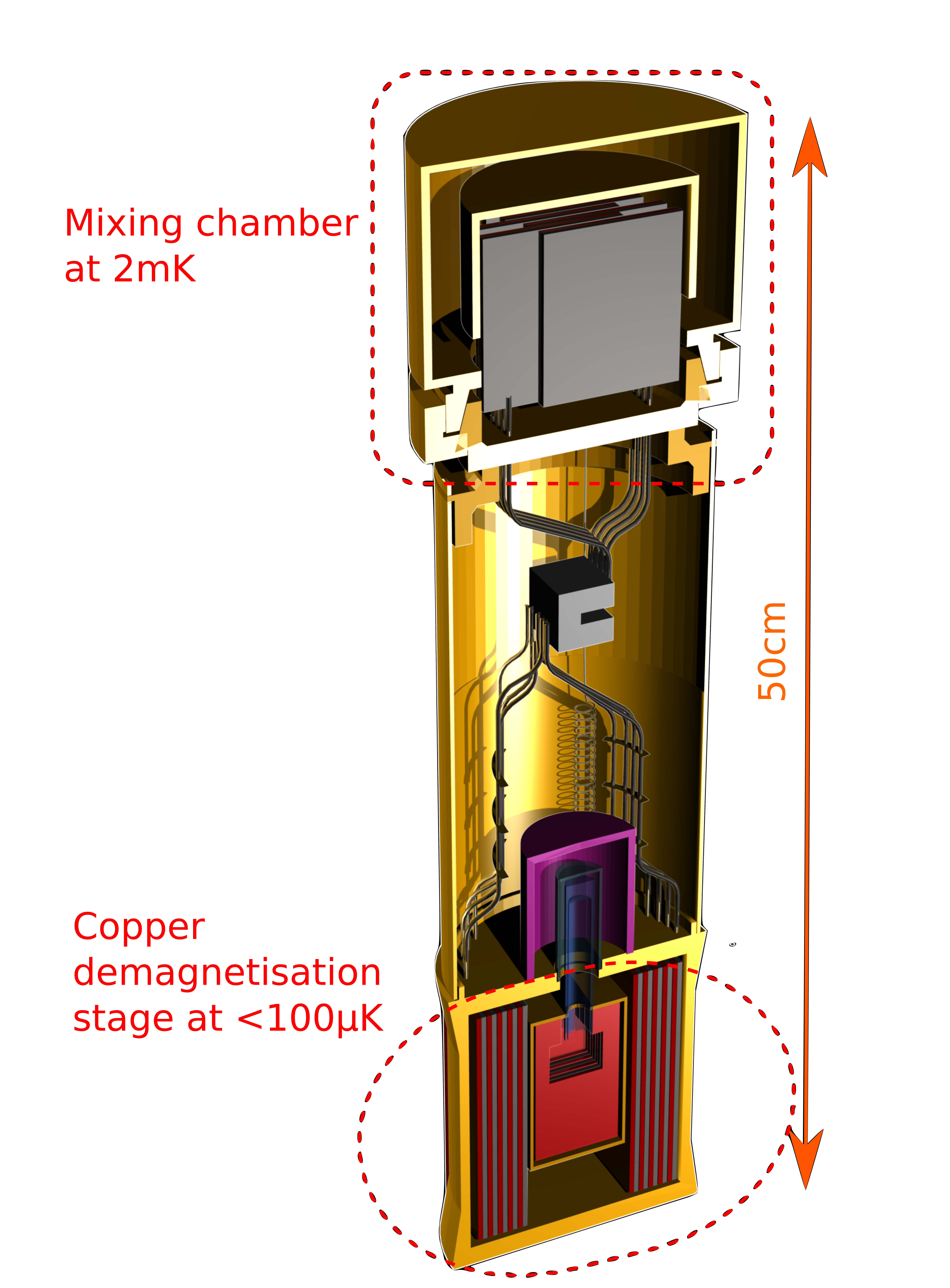}%
    \caption{ Schematic illustration of the QUEST-DMC experiment: The mixing chamber of a $^3$He-$^4$He dilution refrigerator provides a stable temperature of $2$\,mK. This is used to pre-cool a single-shot magnetic cooldown stage made from copper, reaching sub-\SI{100}{\micro\kelvin} temperatures. The superfluid $^3$He dark matter target is contained in a transparent bolometer volume (above the red demagnetisation stage), surrounded by a secondary superfluid volume that is cooled down by the copper coolant. The bolometer is connected to the rest of the superfluid container via an orifice of $\sim1\,\mathrm{mm}^2$ surface area. Scintillation in the bolometer can be monitored by a photon detector (purple shell), but that possibility is not included in the sensitivity analysis presented in this Article. The front half of the whole cylindrical apparatus shown here is cut out and the size of the bolometer is exaggerated for illustrational purposes. The bolometer is instrumented with two resonator wires (not shown here). The magnetic field is vertical.\label{fig:schematic}}
    %On the left is a close-up of the bolometer with }
% is it worth blowing up the cell/purple shell section as a small sub-fig on the left? The detector in this case is the cell/purple shell and is quite small in the figure. -MT   
\end{figure}

The QUEST-DMC detector concept is as follows. In a dark matter-$^3$He scattering event, energy transferred to the struck $^3$He atom will be deposited in the detector volume as heat and ionization energy loss, leading to the formation of thermal excitations (broken Cooper pairs, called quasiparticles) and excimers respectively.
The superfluid $^3$He target is enclosed in a $\sim\!\SI{1}{\centi\metre^3}$ bolometer box, instrumented with a nanomechanical resonator (NEMS) sensitive to thermal quasiparticle production in the $^3$He (see~\cite{autti2023long} for the details of the manufacturing process). The NEMS records the temperature in the box as a function of time~\cite{Pickett_1988,Bradley_1995,hayes1996theoretical}. A sketch of a sample container suitable for such purposes is shown in Figure~\ref{fig:schematic}. To achieve the desired performance, the bolometer is operated at the lowest achievable temperatures, around \SI{100}{\micro\kelvin}; the NEMS is constructed from nanowires with diameters reaching well below the \SI{1}{\micro\metre} scale previously achieved; and, we implement SQUID readout of NEMS at these temperatures for the first time.

In this paper we study the sensitivity of QUEST-DMC using a complete simulation model of the detector and its surroundings. Combining a 
%well-established description
simple model of the superfluid physics, measured NEMS characteristics, radio-assay data of the materials involved, and a comprehensive description of the collision physics using the particle physics simulation library GEANT4~\cite{ALLISON2016186, 1610988, AGOSTINELLI2003250}, we show that the QUEST-DMC detector will provide recoil energy sensitivity down to the $\SI{}{\electronvolt}$ scale. Provided realistic cosmic ray background rejection, this energy sensitivity enables the exploration of the sub-GeV/$c^2$ dark matter mass range with a significant projected improvement in sensitivity to the spin-dependent elastic scattering cross section. 
For example at a dark matter mass of 1 GeV/$c^2$, the leading constraint on the spin-dependent dark matter-neutron cross section stands at $\sim10^{-34}$\,cm$^2$ from CRESST III (LiAlO$_2$)~\cite{CRESST_2022}; the projected sensitivity of a 6-month run of QUEST-DMC with a 4.9\,g\,day exposure in a facility located at the planet's surface reaches down to $\sim10^{-36}$\,cm$^2$.  The details of the profile likelihood ratio sensitivity analysis and background model are described in Sections~\ref{sec:background} and~\ref{sec:sensitivity}.

%The existing leading limit for spin-dependent dark matter-neutron cross section for a dark matter mass of $\SI{1}{\giga\electronvolt \per c^2}$ from CRESST III (LiAlO$_2$)~\cite{CRESST_2022} stands at $\sim10^{-34}$\,cm$^2$; QUEST-DMC is projected to  improve by two orders of magnitude down to $\sim10^{-36}$\,cm$^2$.
% \smw{Not sure we need to state this here as we have plots later in the paper that show it. Suggest we remove it.}
% The current leading constraints in the 0.1-1\,GeV dark matter mass range for spin-independent (-dependent) dark matter scattering interactions is NUMBER  (10$^{-31}$\,cm$^2$)~\cite{CRESST_2022}.  

\section{Detection Principle} 
\label{sec:detection_principle}

\begin{figure}[t!]
 \begin{center}
  \includegraphics[width=0.49\textwidth]{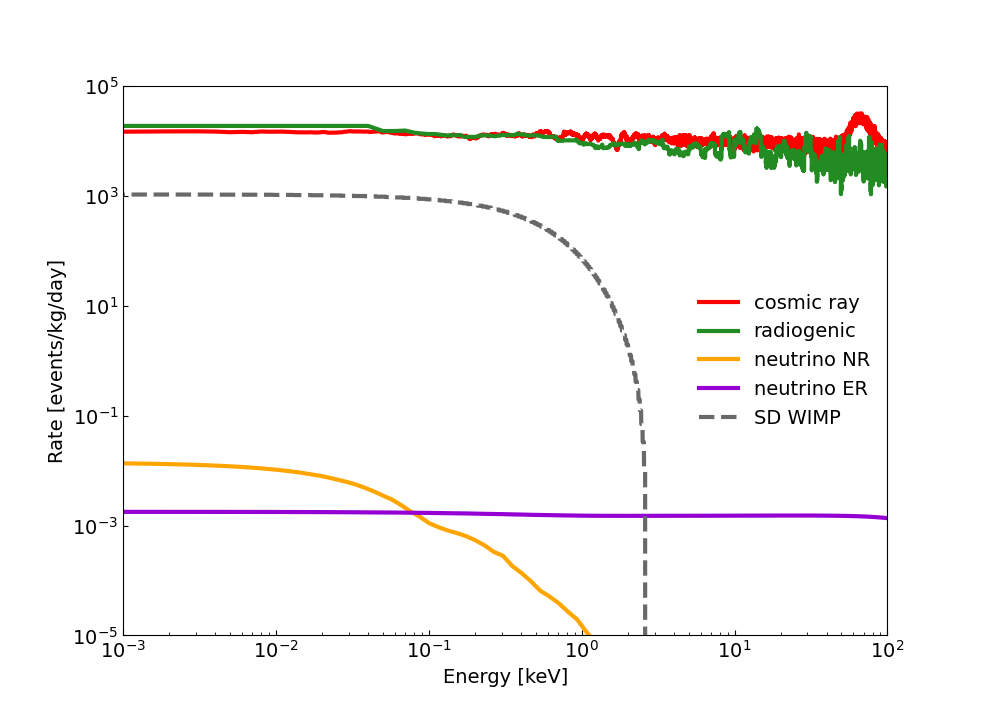} 
 \end{center}
  \caption{Energy spectra showing the results of simulated backgrounds incident on the detector target volume, before the detector response has been applied. As outlined in the Background section, these consist of cosmic rays (including cosmogenics), radiogenic decays of isotopes within the detector materials and surroundings and solar neutrinos - which can result in electron recoil (ER) or nuclear recoil (NR) interactions. The first run of the experiment will take place on the planetary surface with negligible shielding from cosmic rays. A cosmic muon veto system is planned for which a 90\% tagging efficiency is assumed. Here, the nominal radiogenic background for a typical cryostat setup is shown, however this could be further improved using shielding and more radiopure materials. The energy spectrum for spin-dependent dark matter nuclear scattering is shown for comparison using a dashed line, for a dark matter mass of 1\,GeV/$c^2$ and cross section of $10^{-36} \rm cm^2$.}
  \label{fig:bkg_spec}
 \end{figure}

Dark matter can interact with the $^3$He target material via scattering. This may be with the target nucleus, transferring a small fraction of the dark matter kinetic energy, or with the electron cloud, in which case the recoil energy can be as large as the dark matter particle mass. In this paper we focus on nuclear scattering, considering both spin-dependent and spin-independent dark matter interactions. Backgrounds for these dark matter signals include cosmic rays and cosmogenics, radiogenic sources and solar neutrinos. Figure~\ref{fig:bkg_spec} summarises the expected background spectra along with the differential signal event rate for spin-dependent dark matter nuclear scattering of a 1\,GeV/$c^2$ mass dark matter particle.  Details on signal generation and background models can be found in Sections~\ref{sec:signal_generation} and~\ref{sec:background}. For illustration, the maximum recoil energy resulting from interaction of a 1\,GeV/$c^2$ mass dark matter particle is at the scale of 1\,keV.  This drives the detector design of QUEST-DMC to reach the energy threshold of 10\,eV and below.

% In the case of nuclear recoil scattering, which is the focus in this Article, the event rate, $R$ vs. nuclear recoil energy, $E_{\rm NR}$ is given by:
% \begin{equation} \label{eqn:NR_rate}
% \frac{dR}{dE_{\mathrm{NR}}} = \frac{\rho_0 M}{m_N m_\chi} \int_{v_{\mathrm{min}}} v f(\vec{v}) \frac{d \sigma}{dE_{\mathrm{NR}}} d^3v
% \end{equation}
% where $M$ is the detector target mass, $m_N$, is the target nucleus mass, $m_{\chi}$ is the dark matter mass and $\sigma$ is the elastic scattering cross section. Here the astrophysical properties are local dark matter density $\rho_0$ and velocity distribution $f(v)$, which is evaluated in the frame of the detector. 

% The integral is from, $v_{\mathrm{min}}$, the minimum dark matter velocity needed to induce a recoil of energy $E_{\rm NR}$, to $v_{\mathrm{esc}}$, the galactic escape velocity.  As shown in Figure~\ref{fig:bkg_spec}, for a 1\,GeV dark matter particle, the maximum recoil energy is at the scale of 1\,keV.  This drives the detector design of QUEST-DMC to reach the energy threshold of 10\,eV.
%

\subsection{Distribution of collision energy} %Samuli, Dima, Elizabeth
\label{sec:signal_generation}

The energy deposited by a collision is divided between two observable channels, scintillation photons and heat. Heat is ultimately released and detected as superfluid quasiparticles.
We calculate the division of energy by first splitting the initial collision energy into elastic scattering, ionization and excitation. Then each of these is divided into the observable channels and the contributions are summed.

The initial deposited energy, $E$, is split between \linebreak electronic stopping, $\eta$, and nuclear stopping, $\nu$, with \linebreak ${E = \eta(\epsilon) + \nu(\epsilon)}$. For electron recoil (ER) interactions there is little energy transfer to the nucleus and all energy goes into electronic stopping. For nuclear recoil (NR) interactions the fraction of energy in electronic stopping $f$, the nuclear quenching, is calculated using the Lindhard model~\cite{Lindhard}:
\begin{equation}
    f = \frac{\eta(\epsilon)}{E} = \frac{kg(\epsilon)}{1 + k g(\epsilon)}.
\end{equation}
This depends on the atomic number, $Z$, and mass, $A$, through reduced energy, $\epsilon = 11.5 Z^{-7/3} E /\rm keV$ and $k=0.133 Z^{2/3} A^{-1/2}$. The function $g(\epsilon) = 3 \epsilon^{0.15} + 0.7 \epsilon^{0.6} + \epsilon$ is determined from fits to data~\cite{Mei_2008}. At very low recoil energies the Lindhard model is acknowledged to be uncertain, due to the assumed potential and lack of atomic binding energy, this is included in the energy scale systematic described in Section~\ref{sec:sensitivity}

All of the energy in nuclear stopping goes into the elastic scattering channel, but the energy in electronic stopping must be split further into ionization and excitation. This is done using the ratio of measured cross sections for ionization and excitation. For nuclear recoils we use He--He impact cross sections for ionization and excitation (summed over all states)~\cite{ItoSeidel}. No data exists below 1\,keV, so the ratio is extrapolated down to the ionization threshold. For electron recoils we use electron impact cross sections for ionization and excitation to both singlet and triplet states~\cite{ER_datatable}.
% JRM -- added the sentence at the end of this section to address this

The next step is to map the energy deposited in elastic scattering, ionization and excitation, to the quasiparticles and photons seen in the detector. For the elastic scattering all energy goes into quasiparticle excitations. For ionization and excitation channels energy is split between singlet, triplet and infrared photons and quasiparticles, with the ratio depending on both recoil energy and interaction type.

Following ionization, the ejected electron forms a `bubble', repelling nearby He atoms, and the ion forms a `snowball', attracting nearby He atoms. This phenomenon slows the movement of the charges and at small or zero applied field almost all recombine forming excited He$_2^*$ dimers. These subsequently relax, via IR cascades, to the first excited state - singlets ($A^1 \Sigma u$) or triplets ($a^3 \Sigma u$). Both excimers decay emitting UV photons, however the singlet decays on a short (ns) timescale whilst the triplet decay should have a much longer (s) timescale. This should prompt and delay UV scintillation photons, correspondingly. At least one experiment has shown that the triplet lifetime is significantly shorter~\cite{zmeev2013observation} in the presence of $^3$He atoms. The lifetime of triplets in superfluid $^3$He, therefore, remains an open question.

In the case of excitation of a helium atom in the nuclear recoil interaction, the excited helium atom will form an excited dimer. Again, this subsequently relaxes via an IR emission cascade to singlet and triplet states, which emit UV photons as above.

The number of singlet/triplet states depends on the interaction type. For nuclear recoil interactions ionization results in a 0.25:0.75 singlet:triplet ratio from recombination due to the availability of states. Excitation results in a 0.86:0.14 singlet:triplet ratio from transition probabilities~\cite{ItoSeidel}. For electron recoil interactions ionization is assumed to give 0.5:0.5 singlet:triplet ratio from geminate recombination, whilst the singlet:triplet ratio for excitation can be determined from cross section measurements.    

%is 4eV the photon energy? this is near-UV, not IR. Maybe I've misunderstood.... -MT
%Apologies this was badly worded - multiple IR photons are produced when highly excited excimers decay in  a radiative cascade, 4eV is the mean energy going into that channel - EL

A further process that must be accounted for is Penning quenching~\cite{ItoSeidel}, which results in non radiative destruction of two excimers producing helium atoms, an ion and an electron. When the ion and electron recombine a new single excimer is formed, giving a net reduction in the number of excimers and UV scintillation photons. The differential equation for the rate of change of excited states around an interaction site can be solved to give the Penning quenching factor~\cite{Ito_2012}. This modifies the fraction of singlet energy going into UV photons and quasiparticles, with the fraction of singlets lost due to bimolecular processes ranging from $\sim$20\% below 10\,keV to $\sim$40\% for few MeV interactions.

\begin{figure}[t!]
 \begin{center}
    \includegraphics[width=0.49\textwidth]{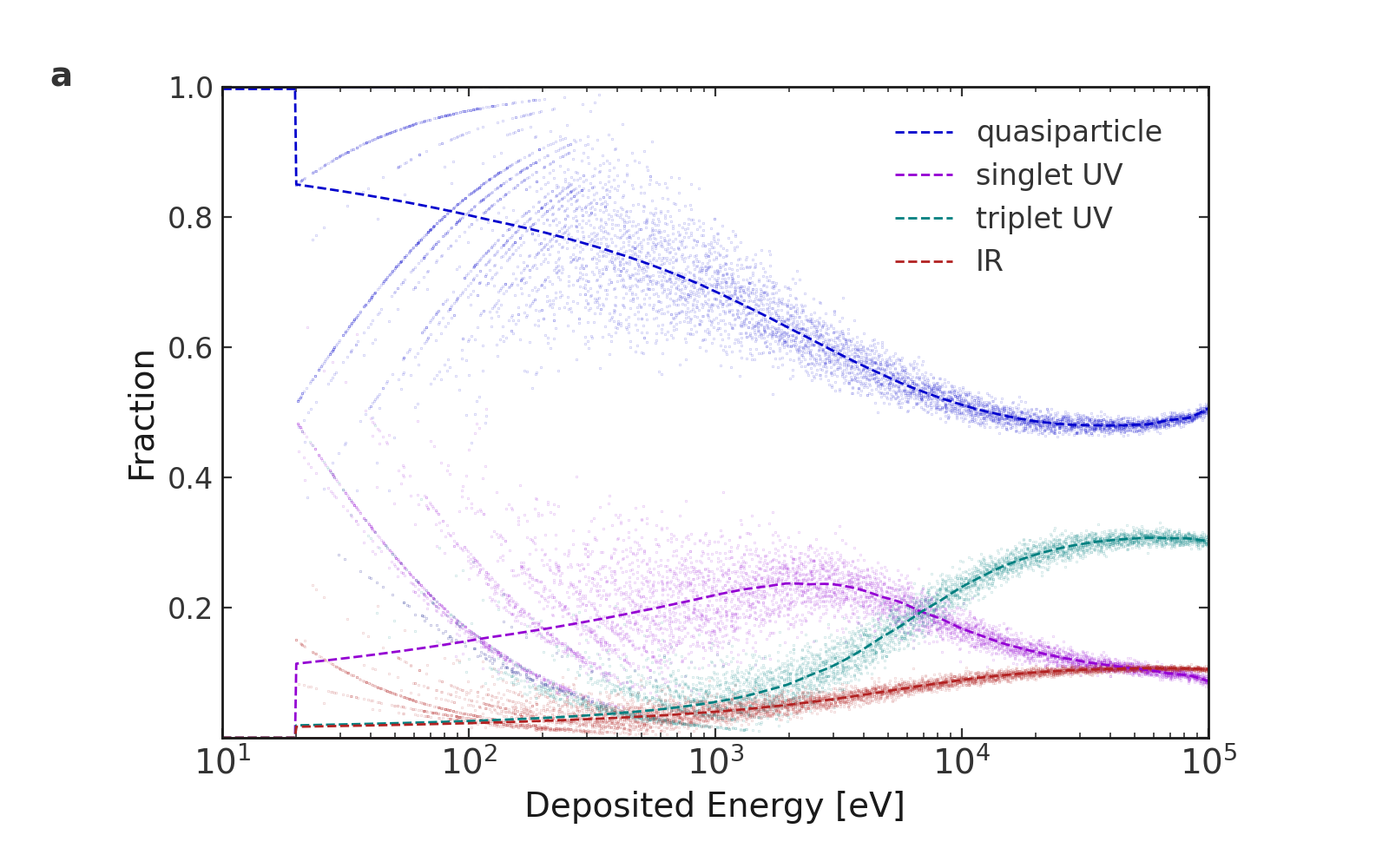} 
    \\[0.1in]
    \includegraphics[width=0.49\textwidth]{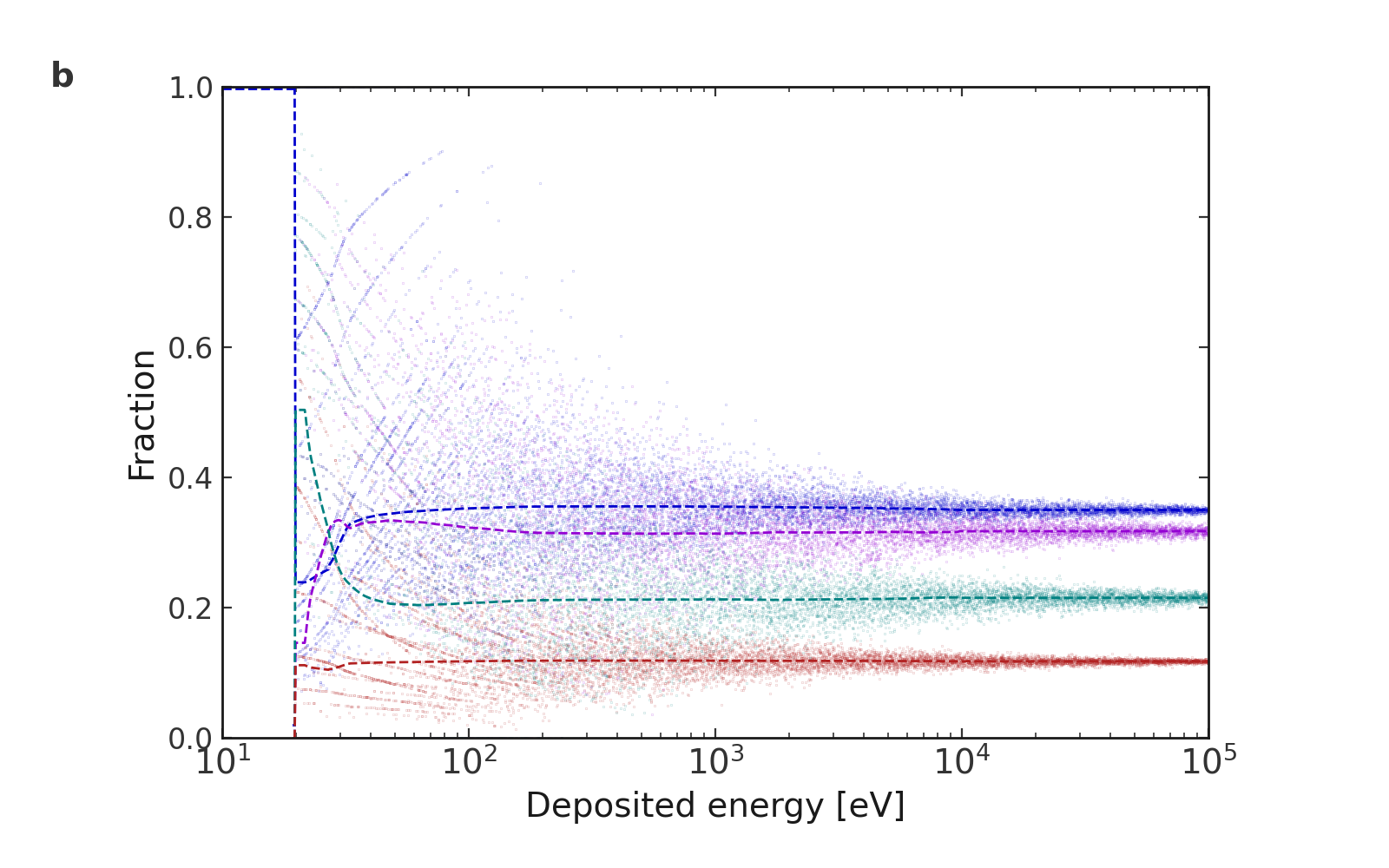}
 \end{center}
  \caption{Partitioning of the deposited energy by (a) nuclear recoil or (b) electron recoil interactions into signal channels. Energy is split between quasiparticles, infrared photons, singlet and triplet UV scintillation photons. Production of quanta is a Poisson process - dashed lines show the mean fraction of energy in each channel and points show Poisson smearing of the discrete quanta produced.
  }
  \label{fig:Edep_partition}
 \end{figure}

%The final step of the energy partition calculation is to sum energies of all quanta. 
%The resulting energy in each channel can be divided by the total deposited energy to give the fraction in each channel. 
The energy going into each channel is summed for each step with quasiparticle energies of 8\,eV from subthreshold electrons, 2\,eV from dimerisation and 4\,eV from ground state dissociation~\cite{ItoSeidel, Seidel_2019Slides}. The mean energies going into the IR channel are 4\,eV from ionization and 0.5\,eV from excitation and mean energies for the UV scintillation photons are 16\,eV for singlets and triplets.
Production of quanta are random processes, which can be modelled using Poisson fluctuations. For each energy, samples are randomly drawn from a Poisson distribution to give a smeared distribution of the number of quanta and the quanta can then be converted back to energy, as shown in Figure~\ref{fig:Edep_partition}.

%Plots

The different fractions of energy in photons and quasiparticles for electron and nuclear recoil events can potentially be used to discriminate between these two interaction types. Below the ionization energy of 19.7\,eV all the energy will go into quasiparticles and for a search below this energy, the photons can be used as a veto for higher energy background events.  The sensitivity projection reported in this paper does not assume any background reduction associated with particle identification based on energy partition, or timing within the ionization partition; however, we note the promise of pulse-shape discrimination in future. 
% NR, ER nuclear recoil and electron recoil? -MT
%Removed abbreviations - EL

\subsection{Detecting the deposited heat}
\label{sec:detection_concept}

\begin{figure*}[htb!]
\centering
  \includegraphics[width=0.90\textwidth]{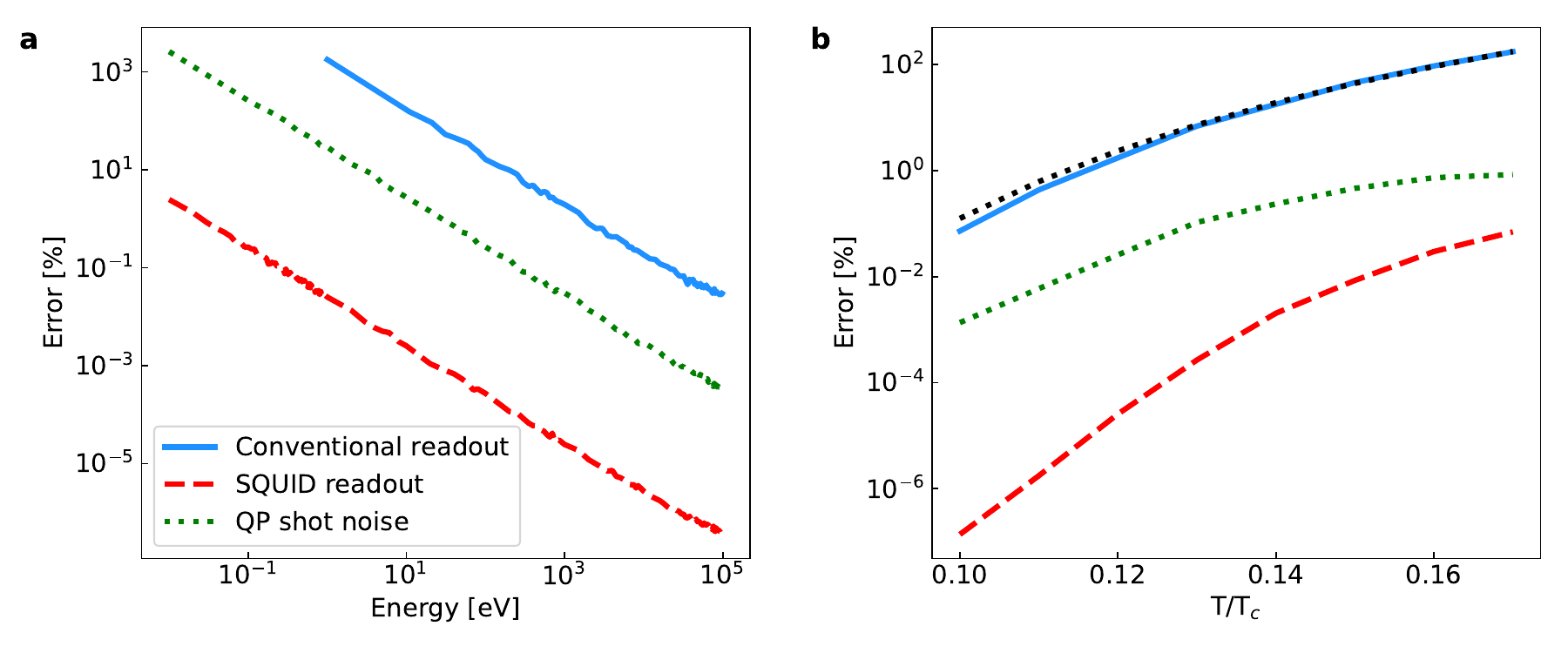}
  \caption{ Simulation of bolometer readout: (a) If the detector is operated at $0.12\,T_\mathrm{c}$ ($T_\mathrm{c} \sim1$\,mK is the superfluid transition temperature), the conventional readout of the thermometer wire (blue line, equivalent amplifier noise $\approx0.1\,\mathrm{nV}/\sqrt{\mathrm{Hz}}$) reaches 10\% uncertainty at deposited energy $Q=100$\,eV. The SQUID readout circuit has equivalent readout noise $\approx0.1\,\mathrm{pV}/\sqrt{\mathrm{Hz}}$, yielding sufficient sensitivity down to sub-eV deposited energies (red dashed line). In both cases the readout error is inversely proportional to the deposited energy $Q$. The shot noise (dotted green line) is shown separately assuming no other noise sources are present. 
  (b) The most important optimisation parameter of the bolometer is the operation temperature. The simulated readout error for fixed deposited energy (here $Q=10$\,eV) is proportional to the thermal quasiparticle density. The black dotted line shows the temperature dependence of the thermal quasiparticle density $\propto\exp(-\frac{\Delta}{k_\mathrm{B} T})$. Both panels show calculations for a 2\,mm-long, 400\,nm-thick detector wire, operated at saturated vapour pressure; the conventional and SQUID readouts are simulated, respectively, for a 100\,mT and 0.4\,mT magnetic field. Dependencies on wire dimensions and operation pressure are weaker than those shown here and are shown in Figure~\ref{fig:error_suppl} in Appendix~\ref{appendix:simulation_of_bolometer}.\label{fig:error}}
\end{figure*}

The detector concept we consider consists of a bolometer cell with a volume of 1\,cm$^3$, filled with superfluid helium-3 in the B phase ($^3$He-B)~\cite{vollhardt2013superfluid}. 
%Thermal excitations in this system (broken Cooper pairs) are called quasiparticles.
In order to maximise the sensitivity of the detector to changes in the quasiparticle density, the ambient quasiparticle density needs to be made as low as possible. This is achieved by cooling the superfluid in the bolometer to $\approx$\SI{100}{\micro\kelvin}. In this temperature regime, thermal quasiparticles propagate ballistically with a mean free path of several kilometres. 

The bolometer is surrounded by large superfluid volume, refrigerated by a nuclear demagnetization refrigerator as shown in Figure~\ref{fig:schematic}. The two volumes are connected via a pinhole in the bolometer wall (diameter $\sim0.5$\,mm), and any deposited heat flows out as carried by the ballistic quasiparticles. This process is described by the bolometer time constant  $\tau_\mathrm{b}\sim5$\,s. The long time constant allows measuring the deposited heat once the collision energy has been thermalised between the quasiparticles in the bolometer (see Appendix~\ref{appendix:superfluid_he3_as_bolometer}). Note that based on the simulations described in the next sections, the expected integrated event rate in the bolometer is $<2.5$ events/minute (without a veto), which is slow as compared with $\tau_\mathrm{b}$. Below we have assumed the bolometer readout cannot be much slower than $\sim10$\,s.

Each quasiparticle carries an energy equal to the superfluid gap $\Delta \approx 10^{-7}$\,eV so that at a recoil energy of 10\,eV, 10$^8$ quasiparticles are created. In comparison, the ionization energy threshold to liberate one electron is $\sim$20\,eV. 

The density of quasiparticles is measured using a nanomechanical wire resonator made from superconducting metal. The moving wire, driven by an AC current in a magnetic field, experiences a drag force proportional to the quasiparticle density. The Full Width at Half Maximum (labelled $\Delta f$) of the mechanical resonance is proportional to the drag force, and the measured signal amplitude to the peak velocity of the probe. In $^3$He-B, this force is orders of magnitude larger than for an equivalent ideal gas~\cite{Cousins_97} because (i) all quasiparticles carry the Fermi momentum instead of a thermal momentum distribution and (ii) quasiparticles approaching the moving probe from behind are Andreev reflected, while those inbound, with a direction of motion opposite to that of the probe, are reflected normally~\cite{bib:Fisher:1989,bradley2017andreev}. Andreev reflection is a quasiparticle reflection process that occurs in quantum condensates that reverses the direction of motion of the quasiparticle but does not transmit momentum (conversion from ``particle'' to ``hole''). Thus, net momentum transfer from the collisions is dramatically amplified.% Note that the quasiparticle density measurement by the nanowire is non-invasive and their number is conserved: quasiparticles are not absorbed like photons or leave the detector volume like rotons and phonons in Refs.~\citenum{HeRALD2019} or~\citenum{DELight2022}.

The quasiparticle drag force measurement is done in two steps:
\begin{enumerate}
  \item In a \textit{sweep measurement} the resonator response is measured as a function of drive frequency to characterise the resonator. The resonance is parametrised by the resonance frequency, the resonance width $\Delta f$, and the amplitude of the response on resonance. These quantities are extracted from fits of a Lorentzian function to the in-phase and out-of-phase part of the voltage. 
  \item If the resonator is driven on resonance, a rapid increase in the quasiparticle drag force produces a decrease in the amplitude of the resonator motion and thus the measured voltage. Once the resonator parameters are extracted using a frequency sweep, driving the resonator on resonance allows extracting such changes as a function of time. This mode of operation is termed \textit{resonance tracking}.
 \end{enumerate}
A particle depositing energy in the bolometer produces an increase in the measured width~\cite{bib:Winkelmann:2007}

\begin{eqnarray}
  \Delta f(t) &= &\Delta f_\mathrm{base}\nonumber + \\ 
  &+& \Delta (\Delta f) {\left( \frac{\tau_\mathrm{b}}{\tau_\mathrm{w}} \right)}^{\tau_\mathrm{w}/(\tau_\mathrm{b}-\tau_\mathrm{w})} \times \nonumber \\ &\times& \frac{\tau_\mathrm{b}}{\tau_\mathrm{b} - \tau_\mathrm{w}} \left( e^{-t/\tau_\mathrm{b}} - e^{-t/\tau_\mathrm{w}} \right).
\label{fit_main}
\end{eqnarray}
Here the energy deposition takes place at time $t=0$ and the peak amplitude $\Delta (\Delta f)$ is proportional to the energy deposited for energies up to several MeV.  The peak shape is determined by the bolometer time constant $\tau_\mathrm{b}$ and the wire time constant $\tau_\mathrm{w}= 1/(\pi \Delta f_\mathrm{base})$, where the latter dependence is valid assuming the change in the width is small as compared with the total width ($\Delta (\Delta f) \ll \Delta f_\mathrm{base}$). Further detail on the underlying physics can be found in Appendix~\ref{appendix:superfluid_he3_as_bolometer}. 

We simulate the bolometer performance by generating bolometer and probe wire responses from Eq.~(\ref{fit_main}). Two techniques of readout have been considered: a passive cold amplifier with amplification factor 100 ("conventional readout"), and a SQUID readout circuit. In both cases, the signal is eventually recorded using a room-temperature lock-in amplifier (see Appendices~\ref{appendix:simulation_of_bolometer} and \ref{appendix:squid}). For the conventional readout the simulated signals are combined with readout noise in the room temperature amplifier. In the case of the SQUID readout circuit the room temperature amplifier noise is negligible, and the noise considered is fundamental noise in the resonator and the cold readout circuit. We fit the obtained noisy signals with Eq.~(\ref{fit_main}), and repeat the process to gather statistics on the error in the fitted outcome. This allows us to calculate the average measurement error as a function of energy deposited by the collision, detector temperature, pressure and other relevant physical parameters. 

The resonance tracking signal measured from the nano\-wire oscillator is much larger than the noise. Thus, signal to noise ratio in detecting bolometer events is directly given by the ratio of $\Delta(\Delta f)$ (in voltage units) and noise. The obtained uncertainty in the recoil energy detection is inversely proportional to the deposited energy (Figure~\ref{fig:error}a). It is also proportional to the quasiparticle density in the bolometer before the collision, which decreases exponentially with decreasing temperature (Figure~\ref{fig:error}b). We conclude that if only readout noise is taken into account, decreasing the temperature in the bolometer is the most effective way of improving the detection sensitivity.

Statistical fluctuations in the quasiparticle collisions with the detector wire %are Poisson distributed, 
result in fluctuations in the measured force proportional to the square root of the measured force (see Appendix~\ref{appendix:shot-noise}). We term this noise quasiparticle shot noise. This noise contribution is shown separately in Figure~\ref{fig:error}. For the detector wire studied here, this noise contribution cannot be observed using traditional readout, but we predict it will be the dominant noise source in a SQUID readout system. Taking the shot noise into account (details in Appendix~\ref{appendix:superfluid_he3_as_bolometer} and~\ref{appendix:simulation_of_bolometer}), we predict that a 10\% readout error can be achieved at deposited energy $Q=10$\,eV using the SQUID circuit up to and possibly even above operation temperature $0.15\,T_\mathrm{c}$.

\begin{figure}[ht!]
 \begin{center}
  \includegraphics[width=0.49\textwidth]{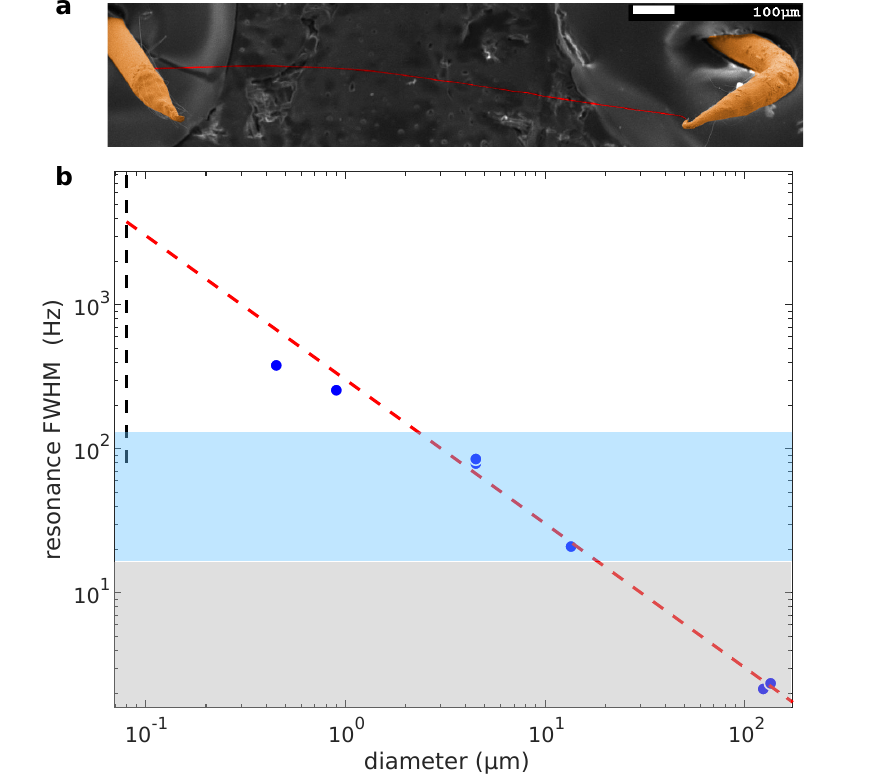}
 \end{center}
  \caption{ Superconducting wire resonators for superfluid thermometry: (a) A fake-colour micrograph of a 390\,nm-thick superconducting wire resonator from above (red string), spanning approximately 1\,mm between two copper posts (copper-coloured rods). The white scale bar corresponds to \SI{100}{\micro\meter}. (b) The blue circles show the Full Width at Half Maximum (FWHM) of thermometer wire resonances for different wire diameters. The data is measured in superfluid $^3$He-B at $0.24T_\mathrm{c}$ (220\,$\mu$K) and saturated vapour pressure (0\,bar). The red dashed line is the theoretical expression detailed in Eq.~(\ref{Eq:Width}) in Appendix~\ref{appendix:superfluid_he3_as_bolometer}. Approaching the coherence length (vertical black dash line), the measured resonance width is expected to remain smaller than the theoretical line owing to qualitative changes in the quasiparticle scattering process. The grey shaded area shows where the resonator response time becomes longer than 10\,s if temperature is lowered to $0.13T_\mathrm{c}$  (\SI{120}{\micro\kelvin} at 0\,bar pressure) and the blue area shows where the response time becomes longer than 10\,s if temperature is lowered to $0.11T_\mathrm{c}$ (\SI{100}{\micro\kelvin}). }
  \label{fig:wire_response_time}
\end{figure}

At the lowest temperatures, the temperature measurement can become too slow to distinguish subsequent collisions. As a crude estimate of the readout speed needed to avoid pile-up, we assume that the detector needs to be able to cope with background events occurring every $\sim10$\,s. The expected background spectra are discussed in detail in the next section. The mechanical resonator adjusts to changes in the quasiparticle density with the time constant $\tau_\mathrm{w} = 1/(\pi \Delta f)$, which increases as

\begin{equation}\label{eq:temp_dep}
\Delta f \propto\exp\left(-\frac{\varDelta}{k_\mathrm{B} T}\right)
\end{equation}
with decreasing temperature. Mechanical resonators also have intrinsic dissipation that does not depend on quasiparticle collisions and thus sets a minimum for $\Delta f$. Lighter resonators (thinner wires) react faster to changes in the quasiparticle density and become saturated at the intrinsic width at a lower temperature. Using sub-\SI{}{\micro\meter} resonators is therefore essential for the operation of the superfluid bolometers.

We have developed a technique of manufacturing superconducting niobium-titanium (Nb-Ti) resonator wires down to 400\,nm thickness. In brief, a copper matrix only containing a few filaments of the desired size is etched away across the planned resonator length, and the filaments are manually removed down to one using tweezers. The one remaining cylindrical superconducting filament remains attached to the copper legs. An example of such resonator~\cite{autti2023long} is shown in the Scanning Electron Microscope picture in the inset of Figure~\ref{fig:wire_response_time}.
%Details of the manufacturing process will be published separately.

Figure~\ref{fig:wire_response_time} shows resonance widths of wires with diameters down to 400\,nm. The shaded areas in Figure~\ref{fig:wire_response_time} indicate an extrapolation to lower temperatures using Eq.~(\ref{eq:temp_dep}), showing which probes become slower than $\tau_\mathrm{w}\approx10$\,s as temperature is lowered to \SI{100}{\micro\kelvin}. That is, probe wires thinner than a micrometre are fast enough to operate in this temperature regime, allowing us to avoid pile-up even if the eV-scale backgrounds turn out to be more frequent than our simulation model predicts. Note that the smallest wires have resonance widths significantly smaller than expected based on theory. This may be either because of impurities attached to the resonator wire in the assembly process or because the wire diameter is approaching the coherence length of the superfluid where Andreev reflection is expected to gradually switch off. For our purposes the obtained widths are large enough for \SI{100}{\micro\kelvin} operation. 

We conclude that the desired energy sensitivity of 10\,eV can be achieved if the $\mathrm{cm}^3$-volume bolometer is operated at \SI{100}{\micro\kelvin} temperature and the readout of a 2\,mm-long thermometer wire is instrumented with a SQUID readout circuit. Achieving sufficient readout speed at this operating temperature to deal with background events requires using a sub-\SI{}{\micro\metre}-thick thermometer wire. 

The bolometer will be energy calibrated using two techniques. Following Ref.~\cite{Fisher_92}, the first technique is to inject heat into the bolometer volume using a second superconducting wire resonator in the bolometer. Intrinsic losses in such a heater are negligible, and therefore the measured driving power is directly dissipated by heating the quasiparticle gas. The steady-state increase in the bolometer temperature is measured as a function of power injected. In this state the power going out via the orifice equals that injected by the heater. The total energy that flows out via the orifice during an event described by Eq.~(\ref{eq:temp_dep}) can be obtained by integrating the corresponding bolometer temperature evolution~\cite{bib:Winkelmann:2007}. The second calibration technique is to place radioactive sources, with fixed energy lines, as near to the active bolometer volume as possible. This calibration gives a direct relation to the deposited energy in the bolometer, from the radioactive decay, to the measured wire response.

\section{Background} % Rob, Paolo, Elizabeth
\label{sec:background}

Beyond reaching low recoil energy sensitivity, the key challenge in dark matter searches is to achieve sufficiently low background event rates in the dark matter search energy region of interest. The $^3$He target itself is intrinsically radiopure.
That is, the only other atomic species that remains liquid at microkelvin temperatures is $^4$He. At \SI{100}{\micro\kelvin}, the solubility of $^4$He in $^3$He is so low that not a single atom is expected to be contained in a bolometer volume of any size. Note also that the rare $^4$He atoms produced by neutron capture will be adsorbed on the container walls after ballistic propagation and that $^4$He atoms in the bulk of the superfluid are not expected to disturb the operation of the bolometer. All other impurities are solidified on filling line surfaces and heat exchangers well before entering the bolometer volume. This extreme purity means that the superfluid has absolutely no intrinsic radioactivity or other background that could impede a dark matter search.
%and that the superfluid is transparent to photons of ultra-violet and visible light. 
As a result, we consider only background sources arising from the materials the detector is made of, as well as external particles incident on the detector--neutrinos and cosmic ray-induced activity.

\subsection{Radiogenic Background Sources} %Rob
% table1: radioassay
% table2: total background rate
% plot with energy distribution?
We calculate estimated rates for radiogenic backgrounds observed in the superfluid detector using a detailed GEANT4 model of the detector geometry and surrounding materials. The model includes the major features of the cryostat and laboratory, and it is based on a combination of dedicated gamma ray spectroscopy assay at Boulby Underground laboratory~\cite{boulby} for all the materials involved that we have direct access to, as well as material radioassay results from the SNOLAB radiopurity database~\cite{LOACH20166}. The specific activities of the main detector materials are summarized in Table~\ref{tab:assay}. 
\begin{table*}[ht!]
\setlength{\tabcolsep}{2.5pt} % reduce the gap between table column
\centering
\caption{{\bf Activity of materials comprising the bolometer, cryostat and surrounding area:} Assay values in mBq kg$^{-1}$ from gamma ray spectroscopy screening \cite{boulby} of materials used in the QUEST-DMC detector, cryostat and laboratory separated into relevant isotopes. * denotes values from dedicated QUEST-DMC screening. Estimate values for materials with no direct access (concrete and insulation of the cryostat's dewar), or isotopic abundances below the sensitivity of dedicated screening, are taken from the SNOLAB radiopurity database~\cite{LOACH20166}. If no value for $^{235}$U could be found the theoretical ratio of natural abundance $^{235}$U/$^{238}$U = 0.007257 is used~\cite{uranium_abundance}. $^{137}$Cs measurement for Araldite epoxy is also used for similar epoxies (Stycast). \label{tab:assay}}
\vspace{0.2 cm}
\begin{tabular}{l|c|c|c|c|c|c|c|c|c|c}
\hline \hline
\rule{0pt}{1.1em}
Material &  Up. $^{238}$U & Lo. $^{238}$U & $^{210}$Pb & Up. $^{232}$Th & Lo. $^{232}$Th & $^{235}$U & $^{137}$Cs & $^{40}$K & $^{60}$Co & $^{54}$Mn\\
\hline 
\rule{0pt}{1.1em}
Concrete & $<1.60\times 10^{5}$ & 1.50$\times 10^{4}$ & $1.00\times 10^{7}$ & 7.57$\times 10^{3}$ & 7.57$\times 10^{3}$ & < 7.20$\times 10^{3}$	& 800 & 4.20$\times 10^{4}$	& < 700 & 0.00\\
Aluminium & 8.33$\times 10^{3}$* & 15.3* & 70.7* & 356* & 334* & 60.5 & < 0.940* & 55.7* & < 1.10* & 0.00*\\
Insulation & 679	& < 200 & < 3.90$\times 10^{3}$	& 200	& 200 & 4.93	& 0.00 & 3.50$\times 10^{3}$ & 400 & 0.00 \\ 
Stainless Steel &  16* & 2.5* & 82.2* & 3.1* & 3.90* & 0.120 & 2.00	& < 6.20*	& < 5.20* & 1.70 \\
Steel & < 12.4 & 12 & 1.20$\times 10^{4}$ & 4.88 & 4.88 & 3.00 & 2.00 & 34.1 & 30.0	& 1.00 \\
Araldite & < 3.60* & < 4.80* & 14.5* & < 3.40* & < 2.20* & 0.0260	& 2.00 & < 25.5*	& 8.00*	& 0.00* \\
Stycast   &  < 10.5* & < 9.50*	& < 14.9* & < 12.8* & < 6.20* & 0.0762*	& 2.00 & < 122*	& 10.0* & 0.00* \\
\hline
\end{tabular}
\end{table*}

The main contributors to radioactive backgrounds in rare-event experiments tend to come from the decay chains of $^{238}$U and $^{232}$Th which have a high abundance in nature, and can produce electrons, alpha particles, photons from de-excitation, or radiogenic neutrons. Secular equilibrium is assumed for isotopes in the Thorium chain unless known otherwise from particular assay results (with equilibrium breaking at $^{228}$Th) and for isotopes in the upper and lower parts of the Uranium decay chain with equilibrium breaking at $^{226}$Ra and $^{210}$Pb. Radon emanation effects are not simulated in initial studies. Other isotopes analysed include $^{235}$U, $^{60}$Co, $^{40}$K, $^{137}$Cs and $^{54}$Mn. If assay results are not able to detect trace amounts of $^{235}$U, the theoretical ratio of natural abundance $^{235}$U/$^{238}$U = 0.007257 is used~\cite{uranium_abundance}.

Radioactive decays from relevant isotopes, shown in Table~\ref{tab:assay}, are simulated originating from each separate volume comprising the GEANT4 model. For each volume between 10$^{5}$ and 10$^{10}$ primary decays are simulated per isotope, dependent on the distance between the volume and the detector. The normalised deposited energy spectra from each decay are summed over all volumes, taking into account the activity and geometric acceptance of the decay products, to provide an estimate rate of ``true'' deposited energy within the $^3$He. The normalised background recoil energy spectra associated with these probabilities can be seen in Figure~\ref{fig:bkg_spec}. The full contributions from radiogenic backgrounds are summarized in Table~\ref{tab:backgrounds}.

Radiogenic backgrounds could be further reduced by the addition of optimized passive shielding.  For example, we have simulated the addition of 10\,cm thick external lead shielding around the cryostat and 2\,cm of radiopure copper within the cryostat itself in GEANT4, which gives $\sim$65\% reduction in the simulated radiogenic background in the Region Of Interest (ROI) (0-10\,keV), compared to the current background model used to assess the sensitivity reach here.

\subsection{Solar Neutrino Backgrounds} %Elizabeth

Neutrinos produced in nuclear processes in the Sun reach the Earth and undergo nuclear recoil or electron recoil interactions in the target.
Coherent neutrino-nucleus scattering mediated by $Z$ exchange produces NRs that can populate the dark matter search energy region of interest, whilst $Z$ and $W$ boson-mediated neutrino-electron elastic scattering produces ERs. 

The incoming neutrino flux depends on the solar model. We use the recommended normalisation values from~\cite{DMStat_2021}. Experimental values are used for $^8$B (SNO~\cite{SNO_2013}) and $^7$Be (Borexino~\cite{Borexino_2019}), whilst theoretical predictions from the high metallicity B16 solar model~\cite{Vinyoles_2017} are used for all other sources.  

The background contribution from solar neutrinos is summarized in Table~\ref{tab:backgrounds}.  Most solar neutrinos are produced in the initial proton-proton fusion step, which dominates both backgrounds at low energies. Neutrinos produced in the later stages by $^7$Be electron capture and $^8$B positron emission extend to higher energies.

\subsection{Cosmic Ray-induced Backgrounds} %Paolo
%Overground experiment
While low background experiments are usually hosted in underground facilities to provide a natural shielding against cosmic radiation, QUEST-DMC will be initially operated at ground level in the ultra-low temperature cryostats at the Lancaster physics department.
In these conditions, the majority of the background will be caused by cosmic rays, and cosmic ray-induced radiation.

The flux of cosmic radiation at ground level has been estimated using the Cosmic-ray Shower Library (CRY)~\cite{cry} particle generator, together with the GEANT4 detector simulation used for the radiogenic background studies. GEANT4 propagates the primary and secondary particles (including cosmogenics production) inside the cryostat materials, accounting for the total deposited energy inside the $^3$He cell.
We do not distinguish ER and NR, and we identify as cosmic ray-induced background anything that produces energy deposition in the ROI; below 19.7\,eV there is no ionization partition so there is no particle identification information.
%For comparison, we also simulate the effect adding 2\,m of water placed on top of the cryostat, suppressing more than 80\% of the incident cosmic rays.
%Tagger
To mitigate the rate of events coming from cosmic rays, we consider using a tagging setup of plastic scintillator planes at the top and at the bottom of the superfluid sample container cell. The light could be carried to the external readout electronic by a light guide. This tagging scheme has an estimated efficiency of 90\% that would permit to reject most of the particles crossing the cell.
The background contribution from cosmic ray-induced backgrounds, with a tagger of 90\% efficiency, is summarized in Table~\ref{tab:backgrounds}.

The assumed cosmic ray flux in the background model at the surface is 0.017/cm$^2$/s. This would be greatly suppressed if operated underground as is typical for rare event search experiments; i.e. at Boulby Underground Laboratory (2805 metres of water equivalent depth) the muon flux has been measured as (4.09$\pm0.15$)$\times 10^{-8}$/cm$^2$/s~\cite{Robinson:2003zj}.

%Background summary table could go here or in next section with discussion of nuisance parameters
\begin{table}[t!] 
\centering
\caption{
{ Background event rates: Expected mean counts for each background component in the 0--10\,keV energy range in units of events per kg per day and per cell per day. The interaction types are indicated by ER for electron recoil and NR for nuclear recoil. The associated uncertainties on the number of counts arise from material screening, flux models and Monte Carlo statistics. \label{tab:backgrounds}}}
\vspace{0.2 cm}
\begin{tabular}{l c c c}
\hline \hline
%\rule{0pt}{1.1em}
Component & \multicolumn{2}{c}{Expected counts [0-10 keV]}  &  Uncertainty \\
& /kg/day & /cell/day & \\
\hline 
%\rule{0pt}{1.1em}
Cosmic ray  & $1.05 \times 10^5$ & $3.31$ & 11 \% \\
Radiogenic ER & $8.31 \times 10^4$ & $2.61$ & 14 \% \\
Solar $\nu$ ER & $1.51 \times 10^{-2}$ & $4.76 \times 10^{-7}$  &  2 \% \\ 
Solar $\nu$ NR & $6.37 \times 10^{-4}$ & $2.01 \times 10^{-9}$ & 2 \% \\
\hline 
%\rule{0pt}{1.1em}
TOTAL   & $1.88 \times 10^5$ & 5.92 \\   
\hline 
\end{tabular}
\end{table}
%Add column for specific activity of each source

\section{Sensitivity analysis} \label{sec:sensitivity} % Elizabeth

This section summarises the signal model; more details are provided in Appendix~\ref{appendix:dm_signal_model}. The differential event rate for a dark matter particle of mass $m_{\chi}$ scattering with a target nucleus of mass, $m_N$, is given by
\begin{equation}
\frac{dR}{dE_{\rm NR}}=\frac{\rho_{\chi}}{m_{\chi}m_N}\int^{\infty}_{v_{\rm min}} \frac{d\sigma}{dE_{\rm NR}}\;v\;f(\vec{v})\;d^3v,
\end{equation}
where $\rho_{\chi}$ is the local dark matter density and $f(\vec{v})$ is the local dark matter velocity distribution in the rest frame of the detector~\cite{Lewin:1995rx,Savage:2006qr} and $v_{\rm min}$ is the minimum dark matter velocity needed to impart a recoil energy of $E_{\rm NR}$ in the detector. The Standard Halo Model is assumed with the dark matter distributed as an isothermal sphere with an isotropic Maxwell Boltzmann velocity distribution truncated at the escape velocity, $v_{\rm esc}$. We use the Halo parameters recommended in~\cite{DMStat_2021}.

% The Standard Halo Model transformed into the rest frame of the detector is used for the velocity distribution. This assumes dark matter is distributed as an isothermal sphere with an isotropic Maxwell Boltzmann velocity distribution, truncated at the escape velocity. Halo parameters $\rho_{\chi}=0.3 \;\rm GeV/c^2$, $v_{\rm esc}= 544 ~\rm km/s$, $v_{\rm sun}=(11.1, 12.2,7.3) ~\rm km/s$ are used, following the recommendations in Reference\cite{DMStat_2021}.

We consider spin-dependent and spin-independent scattering separately. At zero momentum transfer the differential cross sections for spin-dependent and spin-independent scattering can be written, respectively, as
\begin{align}
    \frac{d\sigma^{\rm SD}}{dE_{\rm NR}}
    &=\sigma^{\rm SD}_{\rm \chi n}\frac{m_N}{2\mu^2_{\rm \chi n}v^2}, \\
    \frac{d\sigma^{\rm SI}}{dE_{\rm NR}}&= \sigma^{\rm SI}_{\rm \chi p}\frac{A^2 m_N}{2\mu^2_{\rm \chi p}v^2},
\end{align}
where $A=3$ for $^3$He, $\sigma^{\rm SD}_{\rm \chi n}$ and $\sigma^{\rm SI}_{\rm \chi p}$ are the spin-dependent dark matter-neutron and spin-independent dark matter-nucleon cross sections respectively and where $\mu_{\rm \chi n}$ and $\mu_{\rm \chi p}$ are the reduced dark matter-neutron and dark matter-nucleon masses respectively. 

The differential event rates for spin-dependent and spin-independent scattering are then
\begin{eqnarray}
    \frac{dR^{\rm SD}}{dE_{\rm NR}}=
    \frac{\rho_{\chi}\,\sigma^{\rm SD}_{\rm \chi n}}{2 m_{\chi}\,\mu^2_{\rm \chi n}}
    \int^{\infty}_{v_{\rm min}} \frac{1}{v}\;f(\vec{v})\;d^3v, \\
    \frac{dR^{\rm SI}}{dE_{\rm NR}}=
    \frac{9\rho_{\chi}\sigma^{\rm SI}_{\rm \chi p}}{2 m_{\chi}\,\mu^2_{\rm \chi p}}
    \int^{\infty}_{v_{\rm min}} 
    \frac{1}{v}\;f(\vec{v})\;d^3v.
\end{eqnarray}
 Detector response effects are applied to the predicted differential event rates with energy from the readout error, shot noise and intrinsic fluctuation outlined above and a lower limit determined by the threshold energy of the detector.

%Energy threshold
The energy threshold of the bolometer is calculated as the energy that can be statistically determined to be non-zero at 95\% confidence level. For a range of deposited energy values, the mean number of quasiparticles is calculated as described in Section~\ref{sec:signal_generation}. A smeared distribution about this mean is made using the readout error and intrinsic fluctuations calculated as described in Appendix~\ref{appendix:shot-noise}. Then, for each true number of quasiparticles, a Gaussian is fitted to the smeared distribution. The fraction of this Gaussian which falls below zero is calculated and the true number of quasiparticles corresponding to 5\% below zero is determined. This number of quasiparticles is then converted into an energy, which defines the threshold.

Using this method an energy threshold of 31\,eV is found for the conventional readout and a threshold of 0.51\,eV for the SQUID readout, at a temperature of $0.12\,T_\mathrm{c}$. These threshold energies are used as energy cuts on the background and signal distributions that go into the sensitivity projection.

For the sensitivity projections that follow, an exposure of 4.9\,g\,day is assumed, corresponding to 5 $\times$ 0.03\,g cells with 6 months live data taking time. 

%\subsection{Statistical analysis} %Elizabeth

A profile likelihood ratio analysis is used to evaluate the sensitivity to dark matter interactions~\cite{Cowan_2011,DMStat_2021}. An unbinned likelihood function is defined as the product of Poisson probabilities for the signal plus background rate in all bins. The parameter of interest is defined as the dark matter interaction rate. Other unknown parameters in the model are nuisance parameters, which will be profiled out in the analysis. These include the rates of different background components in the background model, the galactic escape velocity of dark matter in the signal model $v_{\textrm{esc}}$ and the energy scale calibration. 

Using the RooStats package~\cite{RooStats} a test statistic is constructed using the ratio of the conditional to global maximum likelihood. This is evaluated for different values of the parameter of interest, floating all nuisance parameters within Gaussian constraints. The resulting test statistic distribution is used to determine two sided 90\% confidence limit on the parameter of interest, which is converted to a limit on the dark matter interaction cross section.

The probability density functions used in the likelihood function are distributions of the dark matter signal and expected backgrounds in a reconstructed energy variable. The reconstructed energy probability distribution functions are made by applying the energy threshold cutoff and Gaussian energy resolution to ``true'' energy probability distribution functions which have been simulated. Background rates and uncertainties used are shown in Table~\ref{tab:backgrounds}.

The likelihood is extended to include constrain terms on the nuisance parameters accounting for systematics in the signal and background models. For each background component a Gaussian constraint term with width equal to the rate uncertainty is added - arising from solar neutrino flux uncertainty, radiogenic material screening and Monte Carlo simulation statistics.

Uncertainty in the signal model can arise from the astrophysical parameters, which determine the dark matter velocity distribution. Changes in galactic escape velocity can significantly affect the endpoint of this distribution, which is important at low dark matter masses for low threshold experiments. The recommended value of 544\,km/s~\cite{DMStat_2021} was measured by the RAVE survey~\cite{RAVE_2007} with a 90\% confidence limit of (498, 608)\,km/s. The value of $v_{\mathrm{esc}}$ is allowed to float in the analysis, with a Gaussian constraint term corresponding to the 1 $\sigma$ width.

The final source of uncertainty considered in this analysis is in the energy measured in the bolometer - accounted for using an energy scale parameter. This arises from uncertainty in the expected quasiparticle energy, from the deposited energy distribution, transport and cell thermalisation, and from uncertainty in the energy calibration and measurements. The calibration uncertainty is expected to dominate with $\sim$10\%, as seen in the measured calibration coefficient in~\cite{Fisher_92}, and a corresponding Gaussian constraint term is included. 

\begin{figure}[t!]
 \begin{center}
    \includegraphics[width=0.49\textwidth]{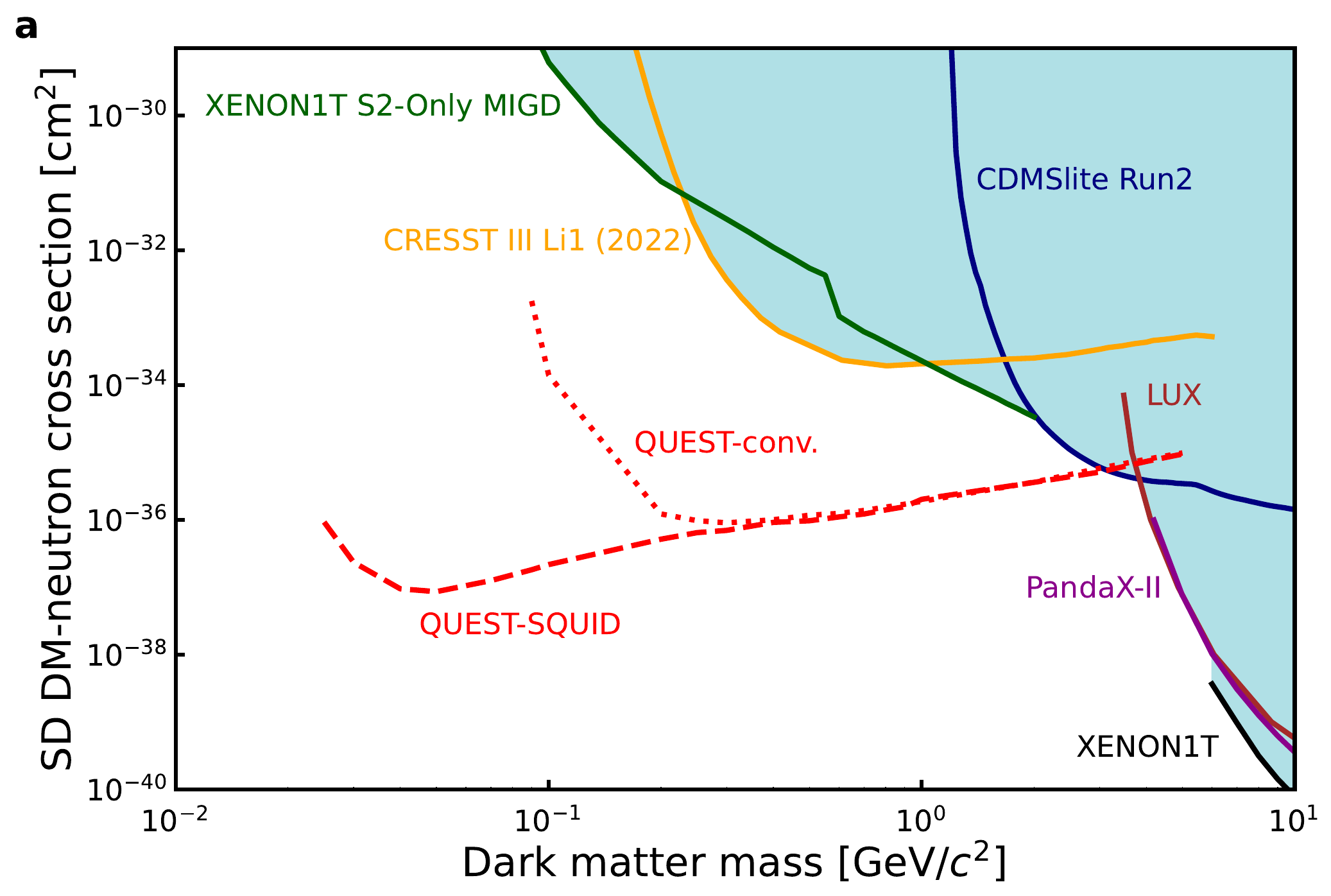} 
    \\[0.1in]
    \includegraphics[width=0.49\textwidth]{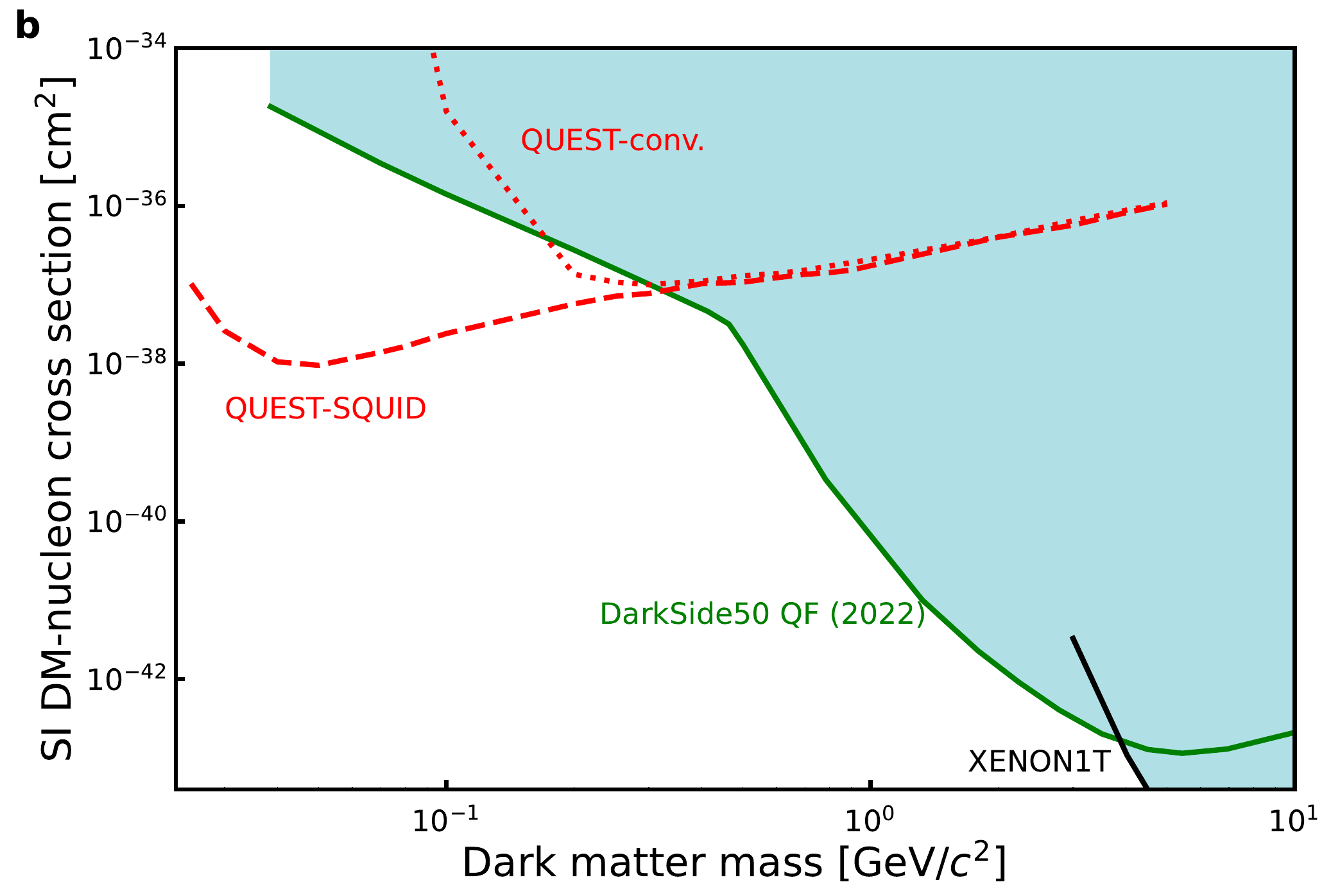}
 \end{center}
  \caption{Projected 90\% exclusion limit sensitivity to dark matter nucleon interactions, assuming 4.9\,g\,day exposure and the background rates outlined in Table~\ref{tab:backgrounds}. Calculated for the conventional (cold transformer plus lock-in amplifier) and SQUID readout schemes, which are expected to achieve energy thresholds of 31 and 0.51\,eV respectively. (a) Cross section limit for spin-dependent dark matter-nucleon interactions including existing limits from Xenon 1T S2-only MIGD~\cite{XENON:2019zpr}, CRESST III (LiAlO$_2$)~\cite{CRESST_2022}, LUX (Xe)~\cite{LUXSD_2016}, CDMSlite (Ge)~\cite{CDMSLite_2018} and PandaX-II~\cite{PandaX-II:2018woa}. (b) Cross section limit for spin-independent dark matter-nucleon interactions including existing limits from DarkSide-50~\cite{DarkSide-50:2022qzh} and XENON1T~\cite{XENON:2018voc}.
  }
  \label{fig:dm_sens}
 \end{figure}

%Sensitivity projection
The resulting 90\% confidence limit sensitivity projection for spin-dependent dark matter-neutron scattering is shown in Figure~\ref{fig:dm_sens} contrasted with existing limits from Xenon 1T S2-only MIGD~\cite{XENON:2019zpr}, CRESST III (LiAlO$_2$)~\cite{CRESST_2022}, LUX (Xe)~\cite{LUXSD_2016}, CDMSlite (Ge)~\cite{CDMSLite_2018} and PandaX-II~\cite{PandaX-II:2018woa}. The projected limits from this work would improve significantly on the existing limits of Xenon 1T~\cite{XENON:2019zpr} and CRESST III (LiAlO$_2$)~\cite{CRESST_2022} in the mass range $\sim(0.025-4)$\,GeV/$c^2$ for the QUEST-DMC SQUID readout. 

The 90\% confidence limit sensitivity projection for spin-independent dark matter-nucleon scattering is also shown in the lower plot of Figure~\ref{fig:dm_sens} along with existing limits from DarkSide-50~\cite{DarkSide-50:2022qzh} and XENON1T~\cite{XENON:2018voc}. Even though $^3$He has a significantly lower atomic mass number compared with xenon and argon, the low energy threshold coupled with the scattering kinematics of the $^3$He with dark matter allows QUEST-DMC to project stronger limits than DarkSide-50 and XENON1T for masses below $\sim$350\,MeV/$c^2$, with the SQUID readout projection providing sensitivity down to $\sim$25 MeV/$c^2$. 

There are a number of sensitivity projections in the literature related to planned upgrades of existing direct detection experiments and to new proposed experiments that are in various stages of development. For spin-dependent dark matter-neutron scattering DarkSPHERE~\cite{NEWS-G:2023qwh} has projections for a helium-isobutane gas target over the mass range $\sim(0.05-10)$\, GeV/$c^2$ reaching as low as $\sim5 \times 10^{-38}$\,cm$^2$ at a dark matter mass of $\sim550$\,MeV/$c^2$. The QUEST-DMC SQUID readout limit is the most sensitive projection below $\sim150$\,MeV/$c^2$. 

% is sensitive to spin-dependent neutron interaction with light dark matter in the 0.05--10 GeV mass range, which operates with a 90\%-10\% mixture of helium-isobutane gas (He:i-C$_4$H$_{10}$) for a total target mass of $27.3$\,kg. 

% For spin-independent dark matter-nucleon scattering, the landscape of projections is more populated with several ongoing experiments with sensitivity projections using the cryogenic bolometers (CRESST~\cite{CRESST:2019jnq}, SuperCDMS~\cite{SuperCDMS:2016wui,SuperCDMS:2022kse}, EDELWEISS~\cite{EDELWEISS:2017uga,Lattaud:2022jnq}), the spherical proportional counters (NEWS-G~\cite{NEWS-G:2023qwh}), the CCD-based (DAMIC-MC~\cite{Castello-Mor:2020jhd}), gas and dual-phase time projection chambers (TREX-DM~\cite{Castel:2018gcp}, DARWIN~\cite{Schumann:2015cpa,DARWIN:2016hyl}) and He-based dark matter detection (HeRALD~\cite{HeRALD2019} and DELight~\cite{DELight2022}). 

For spin-independent dark matter-nucleon scattering, the landscape of projections is more populated for the sub-GeV/$c^2$ mass range. For example, at a dark matter mass of $200$\,MeV/$c^2$ DarkSPHERE~\cite{NEWS-G:2023qwh} projects a limit of $\sim5\times 10^{-42}$cm$^2$. Projections from He-based dark matter detection experiments (HeRALD (kg-day exposure)~\cite{HeRALD2019} and DELight~\cite{DELight2022}) are useful comparisons, with projected limits at $200$\,MeV/$c^2$ of $\sim4 \times 10^{-39}$cm$^2$ and $\sim5 \times 10^{-40}$cm$^2$ respectively, although we note that the DELight projection is background free. We also note that HeRALD~\cite{HeRALD2019} include projections for 100\,kg\,year with a 1\,meV energy threshold giving sensitivity to masses $\sim$MeV/$c^2$ with a cross section of $\sim10^{-45}$\,cm$^2$.

% For the SQUID readout, QUEST-DMC has the most stringent projected limit for dark matter masses below $\sim$60\,MeV/$c^2$ for a 

% , SuperCDMS~\cite{SuperCDMS:2016wui,SuperCDMS:2022kse}, EDELWEISS~\cite{EDELWEISS:2017uga,Lattaud:2022jnq}), the spherical proportional counters (NEWS-G~\cite{NEWS-G:2023qwh}), the CCD-based (DAMIC-MC~\cite{Castello-Mor:2020jhd}), gas and dual-phase time projection chambers (TREX-DM~\cite{Castel:2018gcp}, DARWIN~\cite{Schumann:2015cpa,DARWIN:2016hyl}) and He-based dark matter detection (HeRALD~\cite{HeRALD2019} and DELight~\cite{DELight2022}). 

% The projection sensitivity is strongly affected by exposure, threshold, uncertainties and the intensity of background events.
% In a reconcilable exposure of $O$(kg$\cdot$day), 
% HeRALD with 40\,eV threshold and DELight with 20\,eV threshold (and a zero-background projection) reach masses below 1\,GeV$/c^2$ at high cross-section 
% ($\sigma^{\rm SI}\gtrsim 10^{-39}$\,cm$^2$).

% Currently, QUEST-SQUID has provided the best sensitivity to the spin-dependent DM-neutron
% scattering cross-section at $0.025$\,GeV$/c^2$ and at $0.03$\,GeV$/c^2$ to the spin-independent DM-nucleon cross-section.

%Earth shadowing 
At low dark matter masses and relatively high cross sections it is important to consider the effect of dark matter interactions with particles in the Earth and atmosphere~\cite{Collar:1992qc,Collar:1993ss,Hasenbalg:1997hs,Kouvaris:2014lpa,Kouvaris:2015laa,Bernabei:2015nia,Kavanagh:2016pyr}. Scattering of incoming dark matter will alter the number density and velocity distribution at the detector location. This effect causes the signal spectra and resulting sensitivity to be modified or even lost for sufficiently large dark matter scattering cross sections. In this paper we focus only on the limit at low cross sections, where this effect is minimised, and leave the inclusion of these effects to future work.

% some high cutoff cross section and may modify the velocity distribution used to calculate the signal spectra (and therefore sensitivity) at lower cross sections. 

\section{Discussion} %All
In this paper we have described a new concept for a dark matter experiment, employing superfluid 
$^3$He as a detector for interactions of dark matter with mass of order 1 $\SI{}{\giga\electronvolt \per c^2}$ or below.
The detector concept is based on quasiparticle detection in a bolometer cell by a nanomechanical resonator.   We have developed the energy measurement methodology and detector response model, simulated candidate dark matter signals and expected background interactions, and calculated the sensitivity of such a detector.  We project that such a detector can reach sub-eV recoil energy threshold, opening up new windows on the parameter space of spin-independent interactions of light dark matter candidates. Simultaneously, this complete description of the system allows for future investigations of fundamental superfluid physics, such as understanding the elusive homogeneous first order phase transitions between the superfluid A and B phases where external radiation may trigger the transition instead of thermal fluctuations~\cite{schiffer1995nucleation,halperin2021ab}.

The projected energy threshold of the QUEST-DMC detector is limited by \textit{quasiparticle shot noise} in measuring the energy deposited in the detector. Equipping the bolometer with $N$ independent nanowire oscillators~\cite{noble2022producing,ahlstrom2014quasiparticle} would enhance the projected energy sensitivity by a factor $\sqrt{N}$~\cite{Fleischmann2020}. Alternatively, turning the nanowire into a paddle by attaching a light thin membrane such as a graphene flake or PMMA film would increase the number of quasiparticle collisions without making the resonator significantly heavier, thus reducing the shot noise. In principle, one could also use the shot noise measurement as a direct means of thermometry~\cite{Spietz2003}, or switch to non-local thermometry, such as measuring the relaxation rate of a magnon BEC occupying the bolometer volume~\cite{zavjalov2015measurements,magnon_relax,Heikkinen2014,2000_ppd,autti2018}. Implementing one or more of these improvements will allow pushing the sensitivity to sub-eV energies.

%\section{TODO and notes}
% Paper "deadline" end of June 2023 so to be available for extension discussion

%items already sorted out are commented out below

%% Conclusions needs rewriting to contain
% calibration of the bolometer in association with thermalisation times
% Rob: add details on the full RHUL readout chain.
% PTB squid reference
% explain how the SQUID operational point (B, l, d, etc?) is selected
% Energy partition machinery into Methods/appendix? [Elizabeth]
% Quenching factor uncertainty? [Elizabeth]

% - methods contains no technical details on the particle physics simulation techniques used or the radio assay stuff, should this be added?  JRM: no, not unless peer review raises this
%- Everyone should check that the contributions list appropriately reflects their contributions
%- Biblio: should we always be consistent with "A. Author \textit{et al}" ?
% - Elizabeth: DM figures 2 and 5 captions need updating to follow NC style (start with a bold-font "title", followed by a self-contained explanation of everything in the figure including all symbols used

\section*{Data availability}
The data used in this study are available at \url{https://doi.org/10.17637/rh.23574687}.
% DOI with only the datasets. Data in figures will be added as CSV files.

\section*{Code availability}
The simulation codes are available at \url{https://doi.org/10.17637/rh.23261534}.

\section*{Acknowledgements}
This work was funded by UKRI EPSRC and STFC (Grants ST/T006773/1, EP/P024203/1, EP/W015730/1 and \linebreak EP/W028417/1), as well the European Union’s Horizon 2020 Research and Innovation Programme under Grant Agreement no 824109 (European Microkelvin Platform). We thank George Pickett for the materials used to create Figure~\ref{fig:schematic}. We acknowledge M.G. Ward and A. Stokes for their excellent technical support. S.A. acknowledges financial support from the Jenny and Antti Wihuri Foundation. We thank Boulby Underground Laboratory for radio-assaying the materials. M.D.T acknowledges financial support from the Royal Academy of Engineering \linebreak (RF\textbackslash 201819\textbackslash 18\textbackslash 2).

\section*{Author contributions}
The manuscript was written by S.A., P.F., E.L., J.M., S.M.W., and R.S. with contributions from all authors. The nanowire fabrication and characterisation experiments were carried out by S.A., A.J., J.R.P., T.S., M.D.T., L.W., V.V.Z. and D.E.Z. 
%The experimental cell was designed and built by V.V.Z. 
The radio assay of the cryostat components was carried out by S.A., N.D., P.F., A.K., E.L., J.M., R.S. and D.E.Z. The SQUID readout scheme was designed by S.A., A.C., P.J.H., L.V.L, J.S. and D.E.Z. The numerical simulation model was constructed and used by S.A., N.D., P.F., A.K., E.L., J.M., R.S., S.M.W. and D.E.Z. The project was conceived and supervised by S.A., A.C., R.P.H., M.H., S.J.H., J.M., J.M-R., J.S., V.T., S.M.W. and D.E.Z. 
%D.E.Z. suggested the idea of the detector for QUEST-DMC.

\section*{Competing interests}
The authors declare no competing interests.

\bibliographystyle{apsrev4-1}
%\bibliographystyle{naturemag}
%odd command that overleaf recommend. Now refs appear.
\typeout{}
\bibliography{3He_DM}

\clearpage

%============================================

\appendix
\section*{Appendices}

\section{Superfluid $^3$He as a bolometer}
\label{appendix:superfluid_he3_as_bolometer}

The density of thermal quasiparticles in $^3$He-B decreases exponentially with decreasing temperature. Below $0.25 T_\mathrm{c}$ ($T_\mathrm{c} \sim1$\,mK is the superfluid transition temperature), the density is so low that the quasiparticles do not interact with each other. That is, the mean free path exceeds the sample container dimensions by orders of magnitude. Here, the heat capacity of the superfluid contained by the bolometer can be calculated from first principles~\cite{vollhardt2013superfluid},

\begin{equation}\label{heat_capacity}
C_V(T)=2(2\pi)^{1/2}k_BN_0\varDelta\left(\frac{\varDelta}{k_BT}\right)^{3/2}\exp\left(-\frac{\varDelta}{k_BT}\right)
\end{equation}
where $k_B$ is the Boltzmann constant, $N_0$ is BCS density of states in the normal phase at Fermi energy for one spin component, and $\varDelta$ is the superfluid energy gap. 

The density of quasiparticles can be measured using a mechanical resonator~\cite{autti2023long} immersed in the superfluid. Here we consider a cylindrical superconducting resonator wire, driven by an AC current in a magnetic field.  The velocity of the wire can be measured by recording the induced voltage. The resonator wire experiences a drag force due to collisions with quasiparticles. For a cylindrical resonator wire, the drag force per unit length from quasiparticle collisions follows~\cite{enrico_thesis,Enrico95}

\begin{equation}\label{Eq:damping_force}
F = v d \frac{\pi}{4}p^2_\mathrm{F} v_\mathrm{F} N_0 \exp{\left( -\frac{\varDelta}{k_B T} \right)}.
\end{equation}
Here $v$ is the velocity of the wire, $d$ is its diameter, $p_\mathrm{F}$ is the Fermi momentum and $v_\mathrm{F}$ is the Fermi velocity. This linear expression holds up to $v\approx$\SI{1}{\milli\metre\, \second^{-1}}. In this paper we have therefore assumed that the resonator is operated so that the peak velocity is $v=$\SI{1}{\milli\metre\, \second^{-1}} for superfluid at 0\,bar pressure. Increasing the velocity much beyond this not only makes the drag force nonlinear in velocity but also eventually enables direct vortex production and emission of surface-bound quasiparticles~\cite{autti2020fundamental,Autti_2023}. 

The resonance line shape in the linear regime is Lorentzian, characterised by the central frequency $f_0\sim$1\,kHz and the Full Width at Half Maximum

\begin{equation}
\Delta f = \frac{2F}{\pi^2 \rho d^2 v},
\label{Eq:Width}
\end{equation}
where $\rho$ is the mass density of the wire. The resonance width can be measured directly by sweeping the AC drive frequency across $f_0$ and recording the induced voltage. The resonance width is, therefore, a direct measure of the superfluid's temperature.

The induced voltage depends on the geometry of the resonator, but to a good approximation the on-resonance voltage follows

\begin{equation}
    V_0 \approx l B v 
    \label{eqn:eleven}
\end{equation}
for both semi-loop-shaped resonator wires and straight wires. Here $B$ is the applied magnetic field and $l$ is the length of the wire projected to the plane perpendicular to $B$ (length of the wire for a straight wire, leg spacing for a semi-loop). For the state of the art fabrication of the thinnest wires $l \approx 2$\,mm (longer wires result in greater induced Faraday voltages and, therefore, more sensitive measurements).

To reach the lowest possible temperature, the bolometer needs to be placed as close to the demagnetisation stage as possible, as shown in Figure~\ref{fig:schematic}. Thus, the largest magnetic field available for operating the resonator wires is the typical end-of-demagnetisation field $B = 100$\,mT. We have used this field for calculating the sensitivity of the conventional readout. The SQUID readout can possibly be impedance-matched to work in such field, but would ideally work at a significantly lower magnetic field. We have assumed it is operated at 0.4\,mT as detailed in the following Section.

For a fixed drive current, Eqs.~(\ref{Eq:Width}, \ref{eqn:eleven}) show the product $\Delta f \,V_0$ is conserved. It is therefore possible to operate the resonator in a secondary readout mode by constantly driving it at $f_0$. Changes in the measured $V_0$ can be directly converted into corresponding $\Delta f$. In this mode, the temperature readout time is limited by the intrinsic response time of the resonator, $\tau_\mathrm{w}=1/(\pi \Delta f)$.  

The response of a bolometer operated at temperature $T$, corresponding to thermometer wire resonance width $\Delta f_\mathrm{base}$, is modelled as follows. The heat deposited by a collision in the bolometer volume is denoted $Q$. After the collision, the bolometer reaches a new temperature $\Delta f = \Delta f_\mathrm{base} +  \Delta(\Delta f) $. The increase in the resonator width $\Delta(\Delta f)\propto Q$. %Here we assume the bolometer thermalisation to the superfluid bath around it is infinitely slow.
In practical experiments, the bolometer leaks energy to the surrounding superfluid bath via a `weak link' ---  a small opening in the bolometer wall --- which allows the temperature to slowly return to the value before the collision. A typical hole diameter used is $\sim0.5$\,mm yielding a bolometer time constant $\tau_\mathrm{b}\approx 5$\,s. The superfluid is well-decoupled from the container walls by the Kapitza resistance, that is, the thermalisation time into the bolometer walls is $\gg10^5$\,s. On the other hand, localised heat release up to MeV energies is distributed into the superfluid state within a few microseconds~\cite{ruutu1996vortex} and while the quasiparticle thermalisation time in the bolometer is not known precisely, it is of the same order as the time measured in~\cite{ruutu1996vortex} and thus much shorter than $\tau_\mathrm{b}$. This can be confirmed by measuring the Andreev-reflection-based deviation from Eq.~(\ref{Eq:Width}) at resonator peak velocities $v>$\SI{1}{\milli\metre \second^{-1}} as described by Eq.~(17) in~\cite{fisher1991microscopic} (see also~\cite{enrico_thesis}). This deviation arises from the thermal velocity distribution of the bulk quasiparticles. Therefore, the heat deposited by a collision in the bolometer volume, $Q$, can be measured slowly by recording the resulting thermometer response~\cite{bib:Winkelmann:2007}:
\begin{eqnarray}
  \Delta f(t) &= &\Delta f_\mathrm{base}\nonumber + \\ 
  &+& \Delta (\Delta f) {\left( \frac{\tau_\mathrm{b}}{\tau_\mathrm{w}} \right)}^{\tau_\mathrm{w}/(\tau_\mathrm{b}-\tau_\mathrm{w})} \times \nonumber \\ &\times& \frac{\tau_\mathrm{b}}{\tau_\mathrm{b} - \tau_\mathrm{w}} \left( e^{-t/\tau_\mathrm{b}} - e^{-t/\tau_\mathrm{w}} \right).
\label{fit}
\end{eqnarray}
Note that the peak value of this function is independent of $\tau_\mathrm{b}$ and $\tau_\mathrm{w}$, and thus the readout sensitivity does not depend strongly on either time constant.

The dark matter detection concept presented here relies on detecting changes in thermal energy in the bolometer smaller than any existing experiment is sensitive to. 
While the heat capacity and Andreev reflection responses have not been measured in this energy range,
% [how about ULTIMA measurements of $C_V$ that show surface contribution?]
the theoretical description works well because of the lack of quasiparticle interactions, and so no deterioration of the agreement with theory is expected at even lower quasiparticle densities. This justifies extrapolation to lower-than-yet-unexplored energy scales.

%- vortices, produced by rapid cooling after local heating, also decay quickly but slower than the thermalisation otherwise, could be used to differentiate sources of heat

\begin{figure*}[!htb]
\centering
  \includegraphics[width=0.90\textwidth]{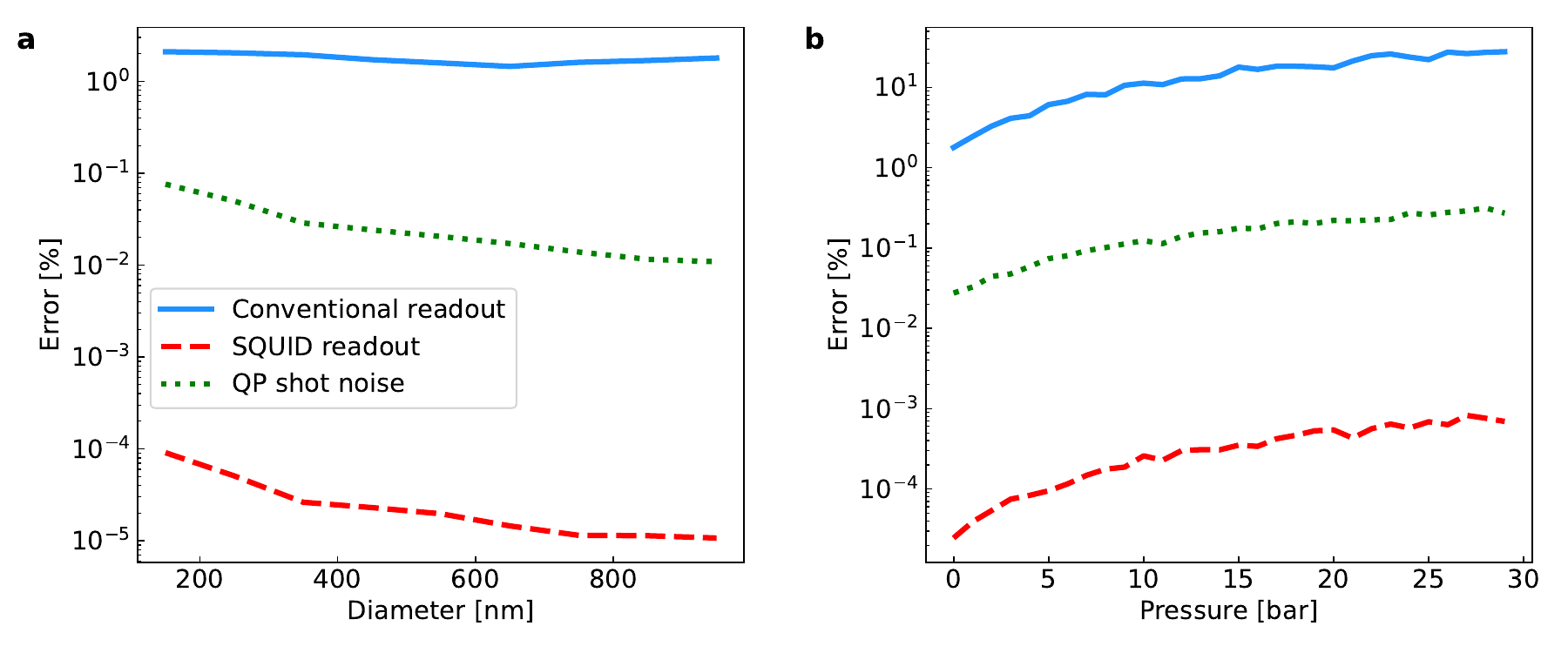}
  \caption{ Simulated bolometer sensitivity as a function of pressure and wire diameter (a) 
  The magnitude of the bolometer response as measured by the thermometer wire only weakly depends on the wire diameter. Thus, the error of the conventional readout (blue line) does not depend on wire diameter. Noise in the SQUID circuit (red dashed line) increases as the wire diameter decreases due to larger thermal motion of the wire, and therefore the readout error increases. The relative shot noise (dotted green line) decreases as the wire size increases due to the increase in the total number of quasiparticles hitting the wire. The pressure in this panel is 0\,bar. (b) The highest sensitivity at a fixed relative temperature ($T/T_\mathrm{c}$) is achieved at zero pressure. Note that the superfluid decouples from heat exchangers at approximately the same $T/T_\mathrm{c}$ regardless of pressure. This panel shows calculations for a 400\,nm resonator wire; 
  the conventional and SQUID readouts are simulated, respectively, for a 100\,mT and 0.4\,mT magnetic field. In both panels the temperature is $T/T_\mathrm{c}=0.12$ and the the collision energy deposited $Q=10$\,eV.}\label{fig:error_suppl}
\end{figure*}

\section{Simulation of bolometer readout}
\label{appendix:simulation_of_bolometer}

We consider two circuits for the readout of induced voltage $V_0$:  a `conventional' four-point measurement with the induced voltage read out via passive pre-amplification by a cold transformer (amplification $\times 100$), and a SQUID-based preamplifier circuit (Figure~\ref{fig:squid}). The latter amplifies the signal orders of magnitude more than the passive transformer, as described below. Both readout schemes use a room-temperature lock-in amplifier to record the signal, operated with time constant 100\,ms. 
%($xxx$\,kHz bandwidth). 
The lock-in input noise is white with the RMS amplitude $V_\mathrm{RMS}$=7.9\,nV. This is the dominant noise source in the passive pre-amplification circuit, while in the SQUID circuit the lock-in noise can be neglected. Noise in the SQUID circuit is derived in the next section.

\begin{figure}[ht!]
 \begin{center}
  \includegraphics[width=0.49\textwidth]{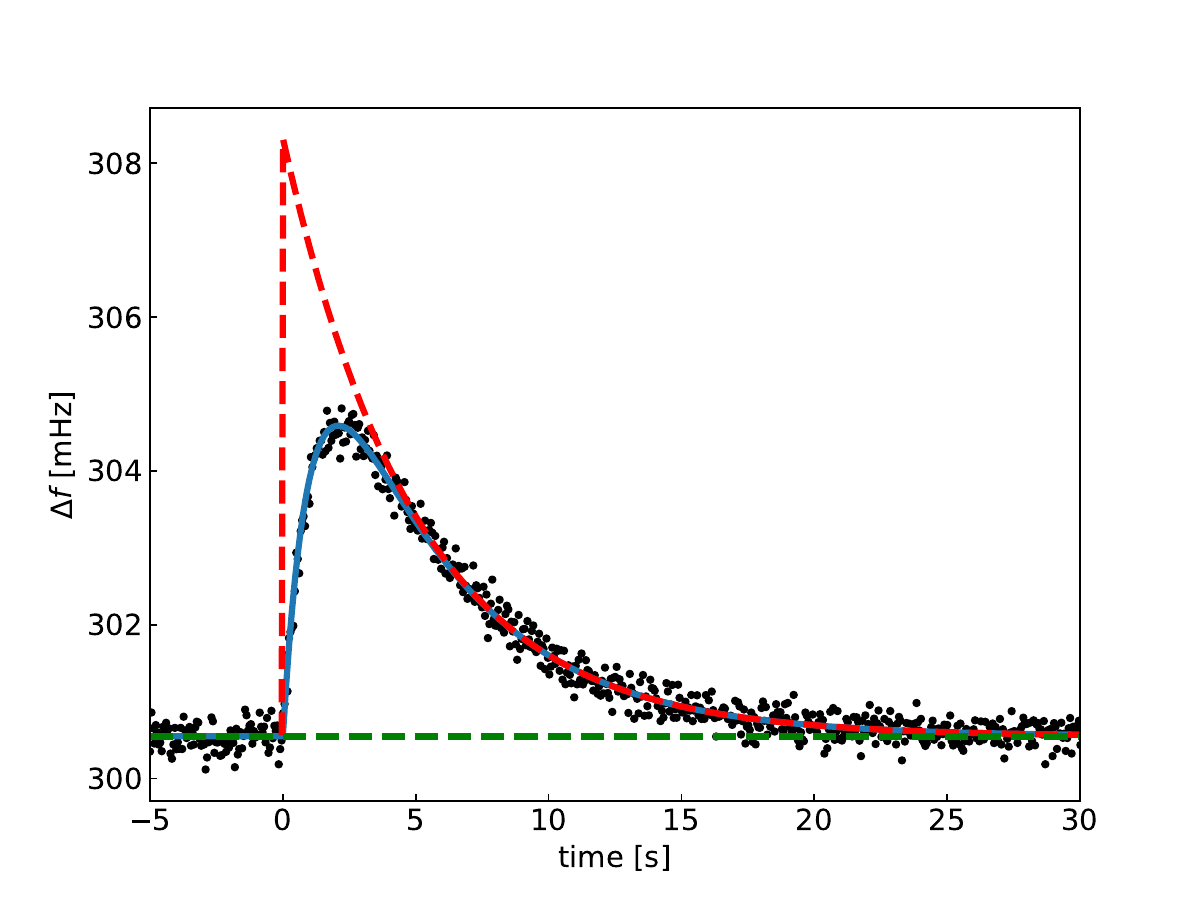}
 \end{center}
  \caption{ A fit to a simulated bolometer response: At $t<0$ the bolometer is at a steady temperature $T/T_\mathrm{c}$ = 0.11. With a 400\,nm readout wire this corresponds to $\Delta f \approx$ 300\,mHz  (black dots). When a sudden heat release of Q = 1\,keV takes place at $t=0$, the bolometer heats up and $\Delta f$ increases as determined by the time constants $\tau_\mathrm{w}$ and $\tau_\mathrm{b}$. The initial temperature rise in the bolometer is assumed to be instantaneous. The blue line shows a fit to Eq.~(\ref{fit}). The bolometer temperature, extracted from the fit, returns to the value before $t=0$ exponentially as determined by $\tau_\mathrm{b}$ (red dash line). The green dashed line shows the base width in the fit.
  Pressure in this simulation is $P=0$\,bar and the noise in the simulated signal corresponds to that in the conventional readout circuit.
  % This would correspond to an uncertainty in the energy measurement of 2\,%.
  %The time alignment is not fitted as for this study the starting time of the energy deposition is assumed to be known.
}
  \label{fig:pulse}
\end{figure}

The resolution of the energy measurement determines the recoil energy threshold in the detector. In order to simulate an experiment, we add noise onto a signal generated using Eq.~(\ref{fit}). The obtained noisy signal is then fitted with the same equation to obtain a fitted deposited energy, and the outcome is compared with the known actual $Q$.
The energy deposition is always considered to take place at $t=0$ so the time alignment is not part of the fitting procedure; this is justified by the fact that, given the current estimated rate of events, any relevant pile up of events is not expected (which would require an additional peak finding procedure).
An example signal with added conventional readout noise is fitted with Eq.~(\ref{fit}) in Figure~\ref{fig:pulse}. Repeating this process yields a Monte Carlo set of pseudo-experiments. The outcome is a Gaussian distribution centred at $Q$ with a standard deviation $\sigma$. The uncertainty shown in Figure~\ref{fig:error} and \ref{fig:error_suppl} corresponds to $\sigma/Q$.

%With the lock-in amplifier readout, 
With the conventional readout scheme, 
the recoil energy can be measured with $<$100\% uncertainty above deposited energy of 100\,eV, whilst with SQUID readout smallest detectable energy with this criterion is approximately 0.1\,eV. The conventional readout sensitivity does not depend on the wire diameter ($\tau_\mathrm{w} \propto 1/d$) as seen in Figure~\ref{fig:error_suppl}a. However, the SQUID readout sensitivity does depend on $d$ because this scheme is sensitive to the thermal motion of the wire as detailed below, and that depends on the thickness $d$. Regardless of the readout scheme, the highest sensitivity is reached at saturated vapour pressure, which is essentially zero at microkelvin temperatures (Figure~\ref{fig:error_suppl}b). We note that helium density increases $\sim$40\% from zero pressure to 30\,bar, and that $T_\mathrm{c}$ is higher at higher pressures so a given relative temperature can be held for longer in practical experiments. These factors need to be taken into account when optimising the detector performance and maximising the useful exposure, but this remains a task for the future.

\section{SQUID readout}
\label{appendix:squid} 

\begin{figure}[ht!]
 \begin{center}
  \includegraphics[width=0.4\textwidth]{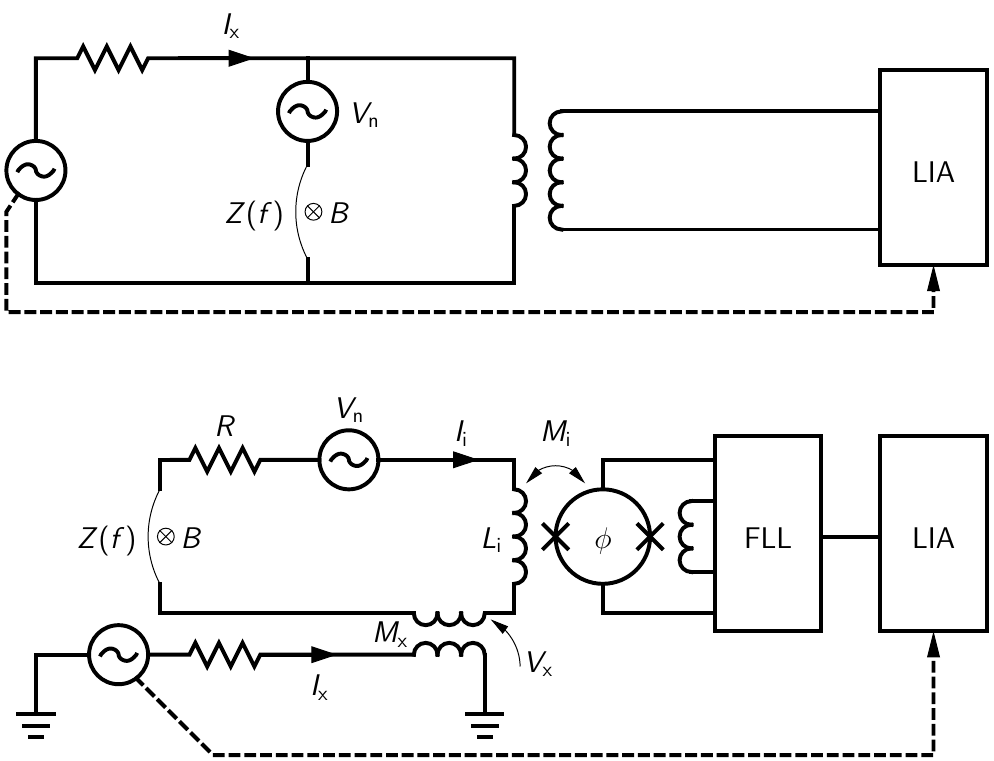}
 \end{center}
  \caption{ Readout schemes for a vibrating wire moving in a magnetic field $B$:
  (Top) Conventional readout scheme. The vibrating wire with impedance $Z(f)$
  is driven with current $I_{\mathrm x}$. The voltage across the wire is detected with a room-temperature lock-in amplifier (LIA), coupled via a cold transformer.
  (Bottom) SQUID readout scheme. The voltage excitation is applied via a transformer with mutual inductance $M_{\mathrm x}$.
  SQUID current sensor detects the current $I_{\mathrm i}$ flowing through the wire,
  contact resistance $R$, and SQUID input coil $L_{\mathrm i}$.
  The SQUID is connected to LIA via room temperature flux-locked-loop (FLL) electronics.
  In both circuits all noise sources can be represented by equivalent voltage $V_{\mathrm n}$ in the wire.
  In our simulation the two readouts are operated at different optimal magnetic fields.}
  \label{fig:squid}
\end{figure}

To improve the energy resolution, we consider nanowire resonator readout
based on a superconducting quantum interference device
(SQUID) current sensor~\cite{Drung2007Electronics}.
The circuit diagram is shown in Figure~\ref{fig:squid}.
The nanowire is excited with a voltage
$V_{\mathrm x} = 2i \pi f M_{\mathrm x} I_{\mathrm x}$
applied by driving a current $I_{\mathrm x}$
via the transformer with mutual inductance $M_{\mathrm x}$.
The impedance of the resonator
\begin{equation}
Z(f) = V_{\mathrm x} \big/ I_{\mathrm i} - R - 2i \pi f L_{\mathrm i}
\end{equation} is inferred from the current $I_{\mathrm i}$ measured by the SQUID,
after subtracting the contact resistance $R$ (the only resistive element of the circuit, we consider an upper bound $R = 1\,\mathrm{\Omega}$)
and impedance of the SQUID input coil $L_{\mathrm i} = 1.5\,\mu$H,
which are connected in series with the nanowire.
The self-inductance of the secondary of the drive transformer is included in $L_{\mathrm i}$.
On resonance the wire impedance $Z_0=Z(f_0)= l B^2 \big/ \big(2 \pi m \, \Delta f \big)$
is directly related to the linewidth $\Delta f$.
Here $m$ is the wire mass per unit length.

We now turn to the sensitivity of the on-resonance readout mode.
We assume the measurement bandwidth $\delta f$ determined by the lock-in time constant
to be greater than $\Delta f$.
It is convenient to represent all noise sources in terms of equivalent voltage
in the wire, shown by $V_n$ in Figure~\ref{fig:squid}.

The intrinsic SQUID flux noise has typical power spectral density
$S_\phi \sim10^{-13} \Phi_0^2/\mathrm{Hz}$, where
$\Phi_0=h/(2 e)\approx 2\,\SI{}{\femto\weber}$ represents the superconducting magnetic flux quantum. This noise corresponds to the input current variance $I_{\mathrm i}^2 = S_\phi \,\delta f / M_{\mathrm i}^2$,
where $M_{\mathrm i} = 10$\,nH is the mutual inductance between the input coil and the SQUID.
The equivalent voltage noise is
\begin{equation}\label{eq:Vn:SQUID}
    V_{\text{SQUID}}^2=\left|Z + R + i 2 \pi f_0 L_{\mathrm i}\right|^2 S_\phi \,\delta f
    \big/ M_{\mathrm i}^2.
\end{equation}
The back action from the SQUID~\cite{Tesche1977squidnoise}
adds a prefactor~\cite{levitin2007tunednmr}
$1 - \alpha^2\eta \sim0.6$ of order unity in front of $L_{\mathrm i}$ in Eq.~(\ref{eq:Vn:SQUID}),
that has no significant effect.
The Johnson-Nyquist noise in the wire contact resistance is
\begin{equation}
    V_{\mathrm{JN}}^2=4 k_B T R \,\delta f.
\end{equation}
The thermal motion of the wire (resonant phonons) leads to
\begin{equation}
V_{\text{wire}}^2=k_B T l B^2 \big/ m.
\end{equation}
Thus, the total RMS noise in the voltage measurement is
\begin{equation}
 V_{\mathrm n} = \small  \sqrt{V_{\text{SQUID}}^2 + V_{\mathrm{JN}}^2 + V_{\text{wire}}^2}.
\end{equation}
Here we assume that the resonant phonons in the wire and electrons in the contacts are at the same temperature $T$ as the surrounding superfluid.

The optimum readout magnetic field is found by comparing $V_{\mathrm n}$ to the voltage $V_0$ associated with the motion of the wire. %Eq.~\eqref{eqn:eleven}.
For the $l = 2$\,mm wire resonating at $f_0 = 0.8$\,kHz
with $\Delta f = 60$\,mHz, considered in our simulations, the best sensitivity is achieved at $B = 0.4$\,mT, where the impedance $Z_0$ of the wire matches the rest of the circuit $R + 2i\pi f L_{\mathrm i}$.
Integrating the SQUID readout into the experimental setup shown in Figure~\ref{fig:schematic} may require tailoring the magnetic field profile or increasing $L_{\mathrm i}$.

%By comparing $V_{\mathrm n}$ to the voltage $V_0$ associated with the motion of the wire, %Eq.~\eqref{eqn:eleven}, we find the optimum readout sensitivity to be achieved in magnetic fields of %order mT, where the impedance $Z_0$ of the wire matches that of the rest of the circuit $R + 2i\pi f %L_{\mathrm i}$.
%The parameters used in the SQUID readout simulation are reported in Table~\ref{tab:squid}.
%\begin{table}[ht!]
%\setlength{\tabcolsep}{2.5pt} % reduce the gap between table column
%\centering
%\caption{Parameters used in the SQUID readout simulation.\label{tab:squid}}
%\vspace{0.2 cm}
%\begin{tabular}{l c c}
%\hline \hline
%Quantity & Value  &  Unit  \\
%\hline 
%$B$      & 0.4   & mT    \\
%$R$      & 1     & $\mathrm{\Omega}$   \\
%$L_{\mathrm i}$    & 1.5   & $\mu$H \\
%%\phi_0 & 2.1   & fWb   \\
%$M_{\mathrm i}$    & 10.0  & nH    \\
%\hline 
%\end{tabular}
%\end{table}

\section{Quasiparticle shot noise}\label{appendix:shot-noise} % Paolo
The three-dimensional calculation that yields the drag force from quasiparticle collisions, Eq.~(\ref{Eq:damping_force}), considers all possible quasiparticle scattering trajectories including those where the quasiparticle is scattered multiple times before leaving the vicinity of the wire. The contribution from escape trajectories requiring multiple scatterings is not negligible~\cite{enrico_thesis}. To estimate the fluctuations in the force, Eq.~(\ref{Eq:damping_force}), we nevertheless neglect these scattering trajectories and simply calculate the quasiparticle flux from the quasiparticle number density. 

The temperature-dependent quasiparticle number density $n$ is
%see J. Vonka thesis page 13 and M. Enrico PhD thesis pages 73 onwards

\begin{equation}
%n = \frac{m^* p_F}{\pi \hbar^2}k_B T \exp \left( {-\frac{\Delta}{k_BT}} \right) \\
n = \frac{1}{\pi^2} \left( \frac{2m^*}{\hbar^2} \right) ^{3/2} \sqrt{E_F} k_B T \exp \left( {-\frac{\varDelta}{k_BT}} \right) \\
\end{equation}
where $E_F = p_F^2/(2m^*)$.
As a two-dimensional approximation, a flux $J=\frac{1}{2}nv_Flr$ of quasiparticles of effective mass $m^*$ impacts on a wire of length $l$ and radius $r$, from one direction.
The QPs hitting the wire follow a Poisson distribution in time. That is, the error (input shot noise, or spectral noise) corresponds to the standard deviation $\sqrt{J}/\sqrt{\mathrm{Hz}}$ in a bandwidth $\delta f$ is expected to be~\cite{Bradley2000}

\begin{align}
\sigma_N &= \sqrt{J} = \sqrt{\frac{nv_Flr}{2\delta f}}.
\end{align}
Therefore the relative error (or fractional noise) per $\sqrt{\mathrm{Hz}}$ is
\begin{equation}
  \sqrt{J}/J = \sqrt{\frac{2}{n v_F l r}}.
\end{equation}
The relative shot noise becomes larger at low temperatures. 

We note that the magnitude of the quasiparticle shot noise in $^3$He has not been verified experimentally, and it may deviate from the above estimate due to the complicated three-dimensional scattering processes involved. If necessary, the shot noise can be reduced significantly by modifying the geometry of the oscillator from a solid-metal cylinder to a paddle-like structure. This can be done by attaching a graphene flake or PMMA film to the superconducting nanowire. This way the flux will significantly increase without making the oscillator much heavier. This modification will change the flow around the sensor and might result in reduction of the critical velocity. Further experiments required to verify this geometry in superfluid $^3$He are underway, however they are outside the scope of this paper.

\section{Signal model}
\label{appendix:dm_signal_model}

We are interested in the sensitivity to both spin-dependent and spin-independent scattering. In this work we consider interactions either arising from $\mathcal{O}_4$ (spin-dependent) or $\mathcal{O}_1$ (spin-independent) in the conventions of~\cite{Fitzpatrick:2012ix}. 

The differential event rate for a dark matter particle of mass $m_{\chi}$ scattering with a target nucleus of mass $m_N$ is given by
\begin{equation}
\frac{dR}{dE_{\rm NR}}=\frac{\rho_{\chi}}{m_{\chi}m_N}\int_{v_{\rm min}}^{\infty} \frac{d\sigma}{dE_{\rm NR}}\;v\;f(\vec{v})\;d^3v,
\end{equation}
where $\rho_{\chi}$ is the local dark matter density, $f(\vec{v})$ is the dark matter velocity distribution evaluated in the frame of the detector, truncated at $v=v_{\rm esc}$ with a lower limit $v_{\rm min}=\sqrt{m_NE_{\rm NR}/2\mu^2_{\chi N}}$, which is the minimum velocity required to produce a recoil energy $E_{\rm NR}$ in the detector, where $\mu^2_{\chi N}=m_{\chi}m_{\rm N}/(m_{\chi}+m_{\rm N})$ is the reduced mass of the dark matter-target nucleus system. 

The differential cross section, $\frac{d\sigma}{dE_{\rm NR}}$, has in general contributions from both spin-dependent and spin-independent processes of the form
\begin{equation}
\frac{d\sigma}{dE_{\rm NR}}=\frac{d\sigma^{\rm SD}}{dE_{\rm NR}}+\frac{d\sigma^{\rm SI}}{dE_{\rm NR}}.
\end{equation}
In order to set useful limits that can be compared across experiments and against theory it is assumed that only one type of process is present. 

Focusing first on the spin-dependent only case. The differential cross section can be written as~\cite{Engel:1992bf}
\begin{equation}\label{eq:diffcrossgf}
    \frac{d\sigma^{\rm SD}}{dq^2}=\frac{8G_F^2}{(2J+1)v^2}\;S_A(q),
\end{equation}
where $q^2=2E_{\rm NR}m_N$ and the spin structure function, $S_A(q)$, is given by
\begin{equation}
    S_A(q)=a_0^2S_{00}+a_0a_1S_{01}+a_1^2S_{11},
\end{equation}
where the $S_{ij}$s are the sum of all contributions to the spin-dependent interactions. The parameters $a_0, a_1$ encode the model dependent isoscalar and isovector effective dark matter-nucleon couplings respectively. 

In the limit of zero momentum transfer ($q=0$), which is a very good limit for dark matter scattering with Helium-3, the spin-structure function can be written as,
\begin{eqnarray}
    S_A(0)
    % &=&\frac{(2J+1)(J+1)}{\pi J}\left(a_p\langle {\bf S_p}\rangle +a_n \langle{\bf S_n}\rangle \right)^2\\
    =\frac{(2J+1)(J+1)}{\pi J}\left(a_p\langle{\bf S_p}\rangle +a_n \langle{\bf S_n}\rangle\right)^2, 
\end{eqnarray}
where $\langle{\bf S_p}\rangle$ and $\langle{\bf S_n}\rangle$ are the mean spins of the proton and neutron in the target nucleus, $J$ is the total spin of the target nucleus and $a_p, a_n$ are the effective dark matter-proton and dark matter-neutron couplings, defined by the relations by $a_{0}=a_p+a_n$ and $a_{1}=a_p- a_n$. For $^3$He, $\langle{\bf S_p}\rangle=0$ leaving only dark matter-neutron spin-dependent scattering~\cite{Catena:2015uha}.

The differential cross section can be written in terms of the spin-dependent dark matter-neutron scattering cross section, $\sigma^{\rm SD}_{\rm \chi n}$, as

\begin{equation}
\frac{d\sigma^{\rm SD}}{dq^2}=\frac{\sigma^{\rm SD}_{\rm \chi n}}{3\mu^2_{\chi n}v^2}\frac{J+1}{J}\langle{\bf S_n}\rangle^2.
\end{equation}

Setting $J=1/2$, $\langle{\bf S_n}\rangle=1/2$ and converting back to use recoil energy in the differential cross section, 
\begin{equation}
    \frac{d\sigma^{\rm SD}}{dE_{\rm NR}}
    % &=&\frac{\sigma^{\rm SD}_{\rm \chi n}\pi}{3\pi\mu^2_{\chi n}v^2}\;\frac{(J+1)}{J}\langle {\bf S_n} \rangle^2
    =\sigma^{\rm SD}_{\rm \chi n}\frac{m_N}{2\mu^2_{\chi n}v^2}.
\end{equation}

% \begin{eqnarray}
%     \frac{d\sigma^{\rm SD}}{dq^2}
%     % &=&\frac{\sigma^{\rm SD}_{\rm \chi n}\pi}{3\pi\mu^2_{\chi n}v^2}\;\frac{(J+1)}{J}\langle {\bf S_n} \rangle^2
%     &=&\frac{\sigma^{\rm SD}_{\rm \chi n}\pi}{4\pi\mu^2_{\chi n}v^2}, \\\sigma^{\rm SD}_{\rm \chi n}&=&\frac{24 G_F^2 a_n^2 \mu^2_{\chi n}}{\pi},
% \end{eqnarray}
Substituting this into the differential event rate we arrive at
\begin{equation}
 \frac{dR^{\rm SD}}{dE_{\rm NR}}=
    \frac{\rho_{\chi}\,\sigma^{\rm SD}_{\rm \chi n}}{2 m_{\chi}\,\mu^2_{\rm \chi n}}
    \int^{\infty}_{v_{\rm min}} \frac{1}{v}\;f(\vec{v})\;d^3v.
\end{equation}

For the spin-independent case, the differential cross section is 
\begin{equation}
    \frac{d\sigma^{\rm SI}}{dE_{\rm NR}}=\frac{m_N\sigma_0 F^2(E_{\rm NR})}{2v^2\mu_{\rm \chi N}^2},
\end{equation}
where $F^2(E_{\rm NR})$ is a nuclear form factor accounting for the loss of coherence for large momentum transfer, $\mu_{\chi N}$ is the dark matter-nucleus reduced mass. The parameter $\sigma_0$ contains the dark matter interaction details and can be written as~\cite{Kopp:2009qt}
\begin{equation}
    \sigma_0=\frac{\left(Zf_p+(A-Z)f_n\right)^2}{f_p^2}\frac{\mu^2_{\chi N}}{\mu^2_{\chi p}}\sigma_{\rm \chi p}^{\rm SI},
\end{equation}
where the parameters $f_{p,n}$ are the dark matter-proton/neutron couplings and $\sigma_{\rm \chi p}^{\rm SI}$ is the spin-independent dark matter-proton cross section. We assume $f_p=f_n$ and in the limit of zero momentum transfer, where $F^2(E_{\rm NR})\rightarrow 1$, the differential cross section simplifies to
\begin{equation}
    \frac{d\sigma^{\rm SI}}{dE_{\rm NR}}=\frac{m_N A^2}{2v^2\mu_{\chi p}^2}\sigma_{\rm \chi p}^{\rm SI}.
\end{equation}
Substituting this into the differential event rate and setting $A=3$ gives
\begin{equation}
    \frac{dR^{\rm SI}}{dE_{\rm NR}} =
    \frac{9\rho_{\chi}\sigma^{\rm SI}_{\rm \chi p}}{2 m_{\chi}\,\mu^2_{\rm \chi p}}
    \int^{\infty}_{v_{\rm min}} 
    \frac{1}{v}\;f(\vec{v})\;d^3v.
\end{equation}

\clearpage

%\onecolumngrid
\clearpage

\setcounter{figure}{0}
\renewcommand{\theequation}{S\arabic{equation}}
\renewcommand{\thefigure}{S\arabic{figure}}
\renewcommand{\thetable}{S\arabic{table}}

\setcounter{equation}{0}
\setcounter{table}{0}

%\section{Extended data}

\end{document}